\documentclass{article}[13pt]
 \usepackage{latexsym}
 \usepackage{amsmath}
 \usepackage{amsfonts}

\usepackage[T1]{fontenc}
\usepackage[utf8]{inputenc}

\usepackage[unicode=true]{hyperref}
 \hypersetup{
   pdfstartview={FitH},
   pdffitwindow=false,
   pdfborder={0 0 0},
   colorlinks=true,
   linktoc=page,
   linkcolor=blue,
   urlcolor=blue,
   hypertexnames=false,
   pdfdisplaydoctitle=true,
   pdfduplex=DuplexFlipLongEdge,
 }

\input{amssym.def}
\input{amssym.tex}

\newcommand\init{\setcounter{equation}{0}}

\newtheorem{theoreme}{Theorem }[section]
\newtheorem{proposition}[theoreme]{Proposition}
\newtheorem{lemma}[theoreme]{Lemma}
\newtheorem{definition}[theoreme]{Definition}

\newtheorem{remark}[theoreme]{Remark}

\newtheorem{conjecture}[theoreme]{Conjecture}
\newcommand{\beq}{\begin{equation}}
\newcommand{\eeq}{\end{equation}}

\def\bel{\begin{lemma}}
\def\eel{\end{lemma}}
\def\bet{\begin{theoreme}}
\def\eet{\end{theoreme}}
\def\bed{\begin{definition}}
\def\eed{\end{definition}}
\def\bep{\begin{proposition}}
\def\eep{\end{proposition}}
\def\ber{\begin{remark}}
\def\eer{\end{remark}}

\newcounter{smallarabics}
\newenvironment{arabicenumerate}
{\begin{list}{{\normalfont\textrm{(\arabic{smallarabics})}}}
  {\usecounter{smallarabics}\setlength{\itemindent}{0cm}
   \setlength{\leftmargin}{5ex}\setlength{\labelwidth}{4ex}
   \setlength{\topsep}{0.75\parsep}\setlength{\partopsep}{0ex}
   \setlength{\itemsep}{0ex}}}
{\end{list}}

\newcounter{smallroman}
\newenvironment{romanenumerate}
{\begin{list}{{\normalfont\textrm{(\roman{smallroman})}}}
  {\usecounter{smallroman}\setlength{\itemindent}{0cm}
   \setlength{\leftmargin}{5ex}\setlength{\labelwidth}{4ex}
   \setlength{\topsep}{0.75\parsep}\setlength{\partopsep}{0ex}
   \setlength{\itemsep}{0ex}}}
{\end{list}}

\newcommand{\ben}{\begin{arabicenumerate}}
\newcommand{\een}{\end{arabicenumerate}}

\def\rr{{\mathbb R}}

\def\zz{{\mathbb Z}}
\def\cc{{\mathbb C}}

\def\bbbone{{\mathchoice {\rm 1\mskip-4mu l} {\rm 1\mskip-4mu l}
{\rm 1\mskip-4.5mu l} {\rm 1\mskip-5mu l}}}
\def\one{\bbbone}

\def\oplusal{\mathop{\hbox{\raise 1.5 ex
  \hbox{$\scriptscriptstyle\rm al$}
\kern -0.92 em \hbox{$\oplus$}}}}
\def\otimesal{\mathop{\hbox{\raise 1.5 ex
  \hbox{$\scriptscriptstyle\rm al$}
\kern -0.92 em \hbox{$\otimes$}}}}
\def\Gammal{\hbox{\raise 1.68 ex 
\hbox{$\scriptscriptstyle\rm al$}\kern -0.50 em $\Gamma$}}

\def\bar{\overline}
\def\met{{\rm met}}
\def\res{{\rm res}}
\def\af{{\rm af}}
\def\ren{{\rm ren}}

\def\nat{{\rm nat}}

\def\sp{{\rm sp}}

\def\Span{{\rm Span}}
\def\s{{\rm s}}

\def\i{{\rm i}}

\def\sgn{{\rm sgn}}

\def\Tr{{\rm Tr}}

\def\Ker{{\rm Ker}}

\def\e{{\rm e}}
\def\d{{\rm d}}
\def\Ran{{\rm Ran}}

\def\w{{\rm w}}

\def\n{{\rm n}}
\def\HS{{2}}

\def\cD{{\cal D}}
\def\cH{{\cal  H}}

\def\cY{{\cal Y}}
\def\cW{{\cal W}}

\def\dg{{\rm dg}}
\def\Im{{\rm Im}}
\def\Dom{{\rm Dom}}
\def\slim{{\rm s-}\lim}

\def\12{\tfrac{1}{2}}
\def\32{\tfrac{3}{2}}
\def\52{\tfrac{5}{2}}

\def\qed{$\Box$\medskip}
\def\proof{{\bf Proof.}\ \ }

\def\z{{\rm z}}
\def\t{\#}

\def\en{{\rm en}}

\begin{document}
\title{Bosonic quadratic Hamiltonians}

\author{Jan Dereziński\\Department of Mathematical Methods in
Physics,\\
 Faculty of Physics, University of Warsaw, \\Pasteura 5,\\ 02-093
 Warszawa, Poland,\\email: jan.derezinski@fuw.edu.pl}
\maketitle
\begin{abstract}
  We discuss self-adjoint operators given formally by 
  expressions quadratic
  in bosonic creation and annihilation operators. We give conditions when
  they can be defined as self-adjoint operators,
  possibly after an infinite renormalization. We
  also discuss explicit formulas for their  infimum.

Our main motivation comes from Local Quantum Field Theory, which
furnishes interesting examples of bosonic quadratic Hamiltonians that require
  an infinite renormalization \cite{De}.
\end{abstract}

\section{Introduction}

{\em Quantum bosonic quadratic Hamiltonians}, or {\em bosonic  Bogoliubov
Hamiltonians} are formally given by expressions of the
form
\begin{eqnarray}\label{exa}
\hat H&=&\sum h_{ij} \hat a_i^*\hat a_j +\12\sum g_{ij}\hat a_i^*\hat a_j^*+\12\sum \bar
g_{ij}\hat a_i\hat a_j+ c,
\end{eqnarray}where $h=[h_{ij}]$ is a Hermitian matrix, $g=[g_{ij}]$
is a symmetric matrix, $c$ is an arbitrary real number (possibly, infinite!) and
$\hat a_i^*,\hat a_j$ are the usual bosonic creation/annihilation operators.
  They are often used in quantum field theory to describe
free theories interacting with a given external classical field
\cite{IZ,De}. They
are responsible for the {\em Casimir effect} \cite{IZ}.  Bogoliubov
applied them to  the theory of interacting Bose gas
\cite{Bo}, which justifies the name
 {\em Bogoliubov Hamiltonians}.

Bogoliubov Hamiltonians that are bounded from below are especially useful. Their infimum $E:=\inf \hat H$
is often interesting
physically.

Bogoliubov Hamiltonians have a surprisingly rich mathematical theory. In infinite dimension this theory sometimes involves
interesting pathologies.  For instance,
 $\hat H$  is often ill defined, but one can define its ``infimum'' $E$. In some situations, one needs
 to perform an infinite renormalization in order to define $\hat H$,
 or at least to compute $E$. This is typical for Bogoliubov
 Hamiltonians that are motivated by relativistic  quantum field theory
 \cite{De}. Another example of interesting
mathematics related to Bogoliubov Hamiltonians can be found in a recent paper
\cite{NNS}, which contains a beautiful proof of diagonalizability of
normally ordered Bogoliubov Hamiltonians under essentially optimal
conditions.

Our paper is devoted to a systematic theory of bosonic Bogoliubov
Hamiltonians in an abstract setting.
We do not restrict ourselves to the
 normally ordered case (with $c=0$ in (\ref{exa})). We start from a
 more general  definition saying that a Bogoliubov
 Hamiltonian is the self-adjoint generator of a one-parameter unitary group on a
 bosonic Fock space that
 implements a  symplectic group. There are interesting
 and physically important examples where the normally ordered
 Bogoliubov Hamiltonian is ill defined, whereas  renormalized ones
 exist \cite{De}.

The family of Bogoliubov Hamiltonians given by fixing $h$, $g$  and
varying $c\in\rr$  (\ref{exa}) can be understood as various quantizations of a
single classical quadratic Hamiltonian,
\begin{eqnarray}\label{exa.1}
 H&=&\sum h_{ij} a_i^* a_j +\12\sum g_{ij} a_i^* a_j^*+\12\sum \bar
g_{ij} a_i a_j,
\end{eqnarray}
where $a_i, a_j^*$ are classical (commuting) variables. $c$, which
appears in (\ref{exa}), can be
understood as the ambiguity of quantization due to noncommutativity of $\hat
a_i$, $\hat a_j$.
The most popular choice is probably $c=0$,
 corresponding to the {\em normally (Wick) ordered Hamiltonian}. It will be denoted $\hat
 H^\n$. The choice $c=\frac12\sum_i
 h_{ii}$, which we call the {\em Weyl Bogoliubov Hamiltonian} and denote
 $\hat H^\w$, has its
 advantages as well. In some situations, however, one needs to
 consider other quantizations, where the constant $c$ may turn out to
 be infinite, and can be viewed as a renormalization counterterm.
 We describe one
 particular possibility, which we call $\hat H^\ren$. In the
 language of Feynman diagrams $\hat H^\ren$ corresponds to discarding loops
 of order 2
 or less, which is often implicit in quantum field theory.

We will use the following notation for the infimum of
the three main Bogoliubov Hamiltonians that we discuss:
\beq E^\w:=\inf \hat H^\w,\ \  \ E^\n:=\inf \hat H^\n,\ \ \ E^\ren:=\inf\hat H^\ren.\eeq
In physics the infimum of the Hamiltonian    appears
under various names, eg. vacuum energy, Casimir energy, vacuum
polarization, effective potential.
Physicists often compute the vacuum energy without worrying whether the
corresponding quantum Hamiltonian is well defined as a self-adjoint
operator. Following this philosophy, we may consider $E^\n$ or
$E^\ren$ under conditions that are more general than the conditions
for the existence of the corresponding Hamiltonians.

\subsection{Comparison with literature}

It is not always very easy
to read the literature
on Bogoliubov Hamiltonians
and to compare statements in various
papers. Their authors often use different conventions, terminology and
notations.

Most of these issues disappear when one fixes a basis in the
1-particle space, identifying it with $\cc^m$. Then a Bogoliubov
Hamiltonian is determined by two matrices, $h=[h_{ij}]$ and $g=[g_{ij}]$,
and possibly a number $c$, see (\ref{exa}).

When we want to use a basis independent language, replacing $\cc^m$
by an abstract Hilbert space $\cW$, it is clear how to interpret
$h$---it is a self-adjoint operator on $\cW$. It is less obvious how
to interpret $g$. One possibility is to view $g$ as a symmetric
tensor, that is, an element of $\otimes_\s^2\cW$. Often, however, it
is preferable to view $g$ as an operator from $\cc^m$ to
$\cc^m$. These two $\cc^m$ should be however viewed as two distinct
spaces---one is the complex conjugate of the other, see eg. \cite{De}.
The notion of a
complex conjugate space is somewhat subtle and has a few equivalent
but superficially distinct interpretations, see Subsection
\ref{Complex conjugate space}. Various authors prefer distinct
interpretations, see eg. the
footnote ${}^6$ in Appendix A of \cite{HS}. (Strictly speaking, this
footnote refers to the fermionic case, however the fermionic and
bosonic cases are quite analogous).

When we consider an infinite dimensional space, there are additional
problems: various operators are often unbounded, are not trace
class, or simply do not exist.

Because of these two kinds of problems, our paper is divided into two
parts.
In the first part we assume that the 1-particle space is finite
dimensional and has a fixed orthonormal basis. All operators are
represented by matrices. We do not worry about conceptual subtleties
related to antilinear maps and the complex conjugate space. Infinite renormalization is not needed
and all formulas are valid with no technical restrictions.

In the second part of our paper, the 1-particle space is an abstract
space $\cW$ of any dimension. We follow mostly the conceptual
framework of \cite{DG}. We distinguish between $\cW$ and its complex
conjugate $\bar\cW$. We need to give technical conditions
guaranteeing that various concepts and formulas survive into infinite dimension.

Throughout the paper it is assumed that the reader is familiar with mathematical formalism of 2nd quantization. Properties of the metaplectic representation in the Fock space play an important role, such as the Shale Theorem and formulas for the Bogoliubov implementers (\ref{metap1})
and  (\ref{metap2}). These formulas were known to Friedrichs \cite{F1}, analysed later by Ruisenaars \cite{Ru1,Ru2} and Berezin \cite{Be}. We treat \cite{DG} as the basic reference on this subject, where in particular various questions related to the unboundedness of bosonic creation and annihilation operators are discussed in detail.

Large parts of Sect. \ref{Finite dimensions, basis dependent formalism} is well known. Thm \ref{th1} about diagonalizability of a quadratic Hamiltonian by
a positive symplectic transformation 
is implicitly contained in \cite{DG} (see Thm
11.20 (3) together with Thm 18.5 (3)).
We come back to this issue in the next section, where  an arbitrary dimension introdces additional technical issues.
Note that a similar fact  proven in
\cite{NNS} does not provide a construction of a distinguished diagonalizing operator.

The basic formula for the infimum of a quadratic Hamiltonian comes from
\cite{BD}. However, some of the formulas for the infimum of the normally ordered Hamiltonians, such as (\ref{fs4}), (\ref{fs4a}) and  (\ref{fs5})
seem to be new. In finite dimension they are not so interesting,
however they become quite useful in  infinite dimension.

It seems that the construction of the renormalized Hamiltonians described in
Subsects \ref{Renormalization I} and  \ref{Renormalization II} has never been presented in the literature in the abstract setting. Their importance is  evident  in concrete situations of Quantum Field Theory described in \cite{De}. We give a brief discussion of the examples from QFT at the end of  Introduction. 

Quadratic Hamiltonians in infinite dimensions is a rather technical topic of operator theory. Therefore, we prefer to give a self-contained treatment of this subject. Many results and definitions that we present are new, however at some places we recall proofs contained in the literature.

Note that it would be awkward and restrictive to define
Bogoliubov Hamiltonians in the infinite dimensional context by an
expression of the form (\ref{exa}). Instead, we define them as
self-adjoint generators of
one parameter unitary groups implementing  Bogoliubov
transformations. (In the bosonic context the term ``Bogoliubov
transformations'' is usually meant to denote ``symplectic
transformations''). The abstract approach makes it sometimes difficult to define some objects, since we cannot
refer to a formula of the form (\ref{exa}). Fortunately,
it is
obvious how to define the Weyl Bogoliubov Hamiltonian---as the generator of a group inside the metaplectic group. It is less obvious how to define normally ordered Hamiltonians. The definition that we propose in Subsec. \ref{Bogoliubov Hamiltonians} seems to be new---in particular, it is more general from the definition of \cite{BD}.

Subsect.
\ref{Bogoliubov Hamiltonians} and \ref{Bogoliubov Hamiltonians-1}
give criteria for the existence of various quantizations. In these subsections there is no assumption on the positivity of $h$. On the other hand, most results require the boundedness of $g$. Some results in this part of the paper come from \cite{Be} and \cite{BD}. However, Thm \ref{th2} (1), which gives a convenient criterion for  the implementability of classical dynamics,
seems to be new. It is  useful in the context of examples from QFT discussed below.

In the following subsections we adopt a different set of assumptions. In particular, we assume that $h$
 is positive and $g$ is form bounded wrt $h$ with bound less than $1$. This condition guaratees the positivity and diagonalizability of classical Hamiltonians.

Diagonalization of Bogoliubov Hamiltonians on the quantum level was considered already by
Berezin \cite{Be}, then by Bach and Bru \cite{BB}.
In a recent paper \cite{NNS}, Napi\'{o}rkowski, Phan Thanh Nam and
Solovej gave a new beautiful proof of diagonalizability. In our paper we
repeat some of
the arguments of
\cite{NNS}, describing their result in
Thm \ref{hanuj}, giving
 essentially
optimal conditions for diagonalization.
In distinction to \cite{NNS}, we show that there exists a
distinguished positive symplectic operator diagonalizing a given
Bogoliubov Hamiltonian.

In Thm \ref{th4} we also describe  a construction of normally ordered Bogoliubov
Hamiltonian based on the form techniques (involving the so-called KLMN
Theorem) presented in
\cite{NNS}. This is an important improvement  (even if it sounds
technical) as compared
to the results of \cite{BD}, which were restricted to operator type perturbations.

 These theorems are complemented with new results.  In  Thm \ref{thmw1}, we show that the dynamics generated by the normally ordered Hamiltonian implements the corresponding classical dynamics. On a formal level this therem seems obvious, nevertheless due to the unboundedness of various operators it needs a careful proof. Another new result, easy in finite dimension and rather technical  in the general case is the formula for the ground state energy described in
Thm \ref{nor3}. We also discuss a criterion for the existence of the
Weyl Bogoliubov Hamiltonians in Thm \ref{th3} and for the existence of the renormalized ground state energy in Thm \ref{pio}.

Let us mention some topics that are left out of our paper. We do not
discuss time-dependent Bogoliubov Hamiltonians, the implementability
and the phase of the corresponding scattering operator. This is
interesting, especially in the context of charged relativistic
fields in an external electromagnetic potential. An infinite
renormalization is needed in order to define the vacuum energy.
This topic on a partly heuristic level is discussed in
\cite{De}. Its fermionic counterpart (a Dirac particle in an external
electromagnetic potential) is better known in the literature, see
eg. \cite{DDMS}.


\subsection{Applications to QFT}


Let us first discuss the question of naturalness of the definition of various kinds of Bogoliubov Hamiltonians.

The Weyl Hamiltonian $\hat H^\w$ is the most natural. In fact, it is invariant wrt symplectic transformations, see (\ref{metap}). Unfortunately, it is often ill defined.

The normally ordered Hamiltonian $\hat H^\n$
is naturally defined given a Fock representation.
In particular, this is the case when we have a distinguished positive classical quadratic Hamiltonian which is treated as the ``free'' one.
Then there exists a unique Fock representation where the free Hamiltonian  has can be quantized without any double creation/annihillation operators.
It is usually quantized in the normally ordered form. We will denote it by
$\hat H_0^\n$

Suppose that we are  interested in the ``full''  Hamiltonian, which is quadratic, but more complicated than the free one and
involves an
interaction with  external fields.
We can then ask whether the corresponding classical Hamiltonian can be quantized.
The most straightforward procedure seems to consider  the normally ordereded full quantum Hamiltonian $\hat H^\n$. Then the corresponding 
ground state energy formally equals then the diference of the ``free Weyl ground state energy'' and the ``full Weyl 
ground state energy'' (in typical situations both infinite).

It sometimes happens that  $\hat H^\n$ is ill defined as well. Then we can try to subtract from $\hat H^\n$ another counterterm. As examples from QFT show, the most natural possibility is to subtract the 2nd order contribution in the perturbative expansion, obtaining $\hat H^\ren$ as in Subsect. \ref{Renormalization I} and \ref{Renormalization II}. Below we briefly describe two examples where one needs to perform such a renormalization. These examples are discussed in more detail in \cite{De}.

Consider the neutral massive scalar quantum field $\hat\phi(\vec x)$. Its conjugate field is denoted $\hat\pi(\vec x)$ with the usual equal time commutation relations
\begin{eqnarray}
[\hat\phi(\vec x),\hat\phi(\vec y)\}=[\hat\pi(\vec x),\hat\pi(\vec y)]&=&0,\nonumber
\\{}
[\hat\phi(\vec x),\hat\pi(\vec y)]&=&\i\delta(\vec x-\vec y).
\label{poisson2}\end{eqnarray}
The free Hamiltonian is defined in the standard way:
\beq\hat H_0^\n:=
\int{:}\Big(\frac12\hat\pi^2(\vec x)+\frac12\big(\vec\partial\hat\phi(\vec
x)\big)^2+\frac12m^2\hat\phi^2(x)\Big){:}\d\vec x,\label{hamiu0}\eeq
where the double dots denote the normal ordering.

Suppose that the mass is perturbed by a Schwartz function $\kappa(\vec x)$. One can check that the normally ordered full Hamiltonian does not exist. However, the renormalized Hamiltonian is well-defined (see Chap. III Subsect. C14 of \cite{De}). Formally, it can be written a
\beq\hat H^\ren:=
\int{:}\Big(\frac12\hat\pi^2(\vec x)+\frac12\big(\vec\partial\hat\phi(\vec
x)\big)^2+\frac12(m^2+\kappa(\vec x))\hat\phi^2(\vec x)\Big){:}\d\vec x-E_2,\label{hamiu}\eeq
where the infinite counterterm
$E_2$ is the contribution of loop diagrams with 2 vertices, see 
Subsect. \ref{Renormalization II}.

The next example is more singular. Consider the charged massive scalar quantum field $\hat\psi(\vec x)$, with $\hat\psi^*(\vec x)$ denoting its Hermitian adjoint. The conjugate field will be denoted $\hat\eta(\vec x)$, so that we have the commutation relations
\begin{align}
  [\hat\psi(\vec x),\hat\psi(\vec y)]&=[\hat\psi(\vec x),\hat\eta(\vec y)]=[\hat\eta(\vec x),\hat\eta(\vec y)]=0,\\
[\hat\psi(\vec x),\hat\eta^*(\vec y)]&=
[\hat\psi^*(\vec x),\hat\eta(\vec y)]=\i \delta(\vec x-\vec y).\label{nonva}\end{align}
The free Hamiltonian is of course
\begin{eqnarray*}
\hat H_0^\n&=&
\int
{:}\Bigl(\hat\eta^*(\vec x)\hat\eta(\vec x)+\vec\partial\hat\psi^*(\vec x)\vec\partial\hat\psi(\vec x) +m^2\hat\psi^*(\vec x)\hat\psi(\vec x)\Bigr){:}\d\vec x.
\end{eqnarray*}

Suppose now that we consider an external stationary electromagnetic potential, described by, say, Schwartz functions $(A^0,\vec A)$.
Then the natural candidate for the full Hamiltonian is (see Chap. VI, Subsec. B17 of \cite{De}):
\begin{eqnarray}\nonumber
\hat H^\ren&=&\int\d\vec x
\Bigl(\hat\eta^*(\vec x)\hat\eta(\vec x)
+\i eA_0(\vec x)\bigl(\hat \psi^*(\vec
x)\hat \eta(\vec x)-
\hat\eta^*(\vec x)\hat\psi(\vec x)\bigr)\\
\nonumber
&&+(\partial_i-\i eA_i(\vec x))\hat \psi^*(\vec x)(\partial_i+\i eA_i(\vec x))
\hat \psi(\vec x)
\\
&&+m^2\hat\psi^*(\vec  x)\hat\psi(\vec x)
\Bigr)-E_0-E_1-E_2,
\label{nmun3}\end{eqnarray}
where  $E_0,E_1,E_2$ are infinite counterterms, which come from the expansion described in (\ref{expa1}).

Unfortunately, the classical dynamics is implementable only
if the vector potential $\vec A$ vanishes everywhere.
Therefore, $\hat H^\ren$ is well defined only in this case.
However, the infimum of (\ref{nmun3}), that is $E^\ren$, is a well defined
gauge-invariant number also for nonzero $\vec A$. 

Note that both Hamiltonians  (\ref{hamiu}) and 
(\ref{nmun3}) can be derived from local Lagrangians. Therefore, even
if the models based on these Hamiltonians
do not
  satisfy Haag-Kastler axioms in the strict sense (because of the
  absence of translation invariance), they belong to Local Quantum
  Field Theory: they lead to nets satisfying the Einstein causality,
  and they have bounded from below Hamiltonians. At the same time, all
  of them require 
  an infinite renormalization, typical for computations in
  perturbative Quantum Field Theory.

  The examples \ref{hamiu0} and \ref{nmun3}
  are especially interesting
  in the context of more complicated interacting quantum field theory when, typically, $\kappa$, $A^0$ and $\vec A$ are promoted to the role of a quantum fields. Then $E^\ren$ can be interpreted as
  the value of certain renormalized loop diagrams. In particular, $E^\ren$ of the second example is usually called the {\em vacuum poarization (in scalar QED)}.

{\bf Acknowledgements.} I would like to thank Marcin Napi\'{o}rkowski,
Phan Th\`{a}nh Tham and Jan Philip Solovej for useful discussions.
I gratefully acknowledge financial support of the National Science
Center, Poland, under the grant UMO-2014/15/B/ST1/00126.

\section{Finite dimensions, basis dependent formalism}
\label{Finite dimensions, basis dependent formalism}
\init

Let us first describe the basic theory of  bosonic quadratic Hamiltonians
in finite dimensions, assuming that the one-particle space is $\cc^m$.
Seemingly, our formulas will depend on the choice of the canonical
basis in $\cc^m$. In reality, after an appropriate interpretation, they
are basis independent. This interpretation  will 
be given in the next section, when we discuss an arbitrary dimension.

 Operators on $\cc^m$ will  be identified with matrices. 
If $h=[h_{ij}]$ is a matrix, then $\bar h$, $h^*$ and $h^\#$ will
denote its complex conjugate, hermitian conjugate and transpose.

\subsection{Creation/annihilation operators}

We consider the bosonic Fock space  $\Gamma_\s(\cc^m)$.
 $\hat a_i, \hat a_j^*$ are the standard annihilation and
creation operators on  $\Gamma_\s(\cc^m)$.
$\hat a_i^*$ is the Hermitian conjugate of $\hat a_i$,
\begin{eqnarray*}[\hat a_i,\hat a_j]=[\hat a_i,\hat a_j]&=&0,\\{}
[\hat a_i,\hat a_j^*] &=&\delta_{ij}.\end{eqnarray*}

(We decorate creation/annihilation operators with hats, because we
want to distinguish them from their classical analogs).

We use the more or less standard notation for operators on Fock spaces. In
particular, we use the standard notation $\Gamma(\cdot)$ and
$\d\Gamma(\cdot)$, which will be recalled in Subsection
\ref{Fock spaces}. If $w=[w_i]\in\cc^m$, then the corresponding
creation/annihilation operators are
\begin{eqnarray}
  \hat  a^*(w):=\sum_i w_i\hat a_i^*,&&
  \hat  a(w):=\sum_i \bar w_i\hat a_i.\label{pio1}
\end{eqnarray}
 If $g=[g_{ij}]$ is a symmetric $m\times m$ matrix, then the
 corresponding double creation/annihilation operators are
\begin{eqnarray*}
  \hat  a^*(g):=\sum_{ij} g_{ij}\hat a_i^*\hat a_j^*,&&
  \hat  a(g):=\sum_{ij} \bar g_{ij}\hat a_j\hat a_i.
\end{eqnarray*}

\subsection{Classical phase space}

To specify a linear combination of operators $\hat a_i$, $\hat a_j^*$
we need to choose a vector
$(w,w')\in\cc^m\oplus\cc^m$:
  \beq\hat\phi(w,w'):=\sum_i \hat a_i^*w_{i}+\sum_i \hat a_iw_{i}'\label{pio0}\eeq
 (\ref{pio0}) is self-adjoint iff $\bar w=w'$.
Therefore, it is natural to introduce  
the doubled
space
$\cc^m\oplus\cc^m$ equipped with the {\em complex conjugation}
\beq
J\left[\begin{array}{c}w\\ w'\end{array}\right]
=\left[\begin{array}{c}\bar w'\\ \bar w\end{array}\right]
.\eeq
Vectors left invariant by $J$ have the
form \beq
\left[\begin{array}{c}w\\\bar w\end{array}\right],\ w\in\cc^m.\label{real}\eeq
They form a $2m$-dimensional real subspace of $\cc^m\oplus\cc^m$,
which can be identified with $\rr^{2m}$. (In  what follows, when we
speak of $\rr^{2m}$ we usually
mean the space of vectors of the form (\ref{real})).

Operators on $\cc^m\oplus\cc^m$ that commute with $J$, or equivalently preserve $\rr^{2m}$,
have the form
\beq
R=\left[\begin{array}{cc}p&q\\\bar q&\bar p\end{array}\right],\eeq
and will be called {\em $J$-real}.
Note that if we know the restriction of $R$ to (\ref{real}) then we can
uniquely extend it to a (complex linear) operator on $\cc^m\oplus\cc^m$.

The operator
\beq
S=\left[\begin{array}{cc}\one&0\\0&-\one\end{array}\right].\label{ess}\eeq
determines the commutation relations:
\beq\left[\hat\phi(w_1,w_1')^*,\hat\phi(w_2,w_2')\right]=
(w_1|w_2)-(w_1'|w_2')=\big((w_1,w_1')|S(w_2,w_2')\big).\eeq

Instead of quantum operators $\hat a_i^*$ and $\hat a_j$, one can
also consider  classical (commuting) variables 
 $a_i$, $a_j^*$, $i=1,\dots,m,$ 
 such that $a_i^*$ is the complex
conjugate of $a_i$ and the following Poisson bracket relations hold:
\begin{eqnarray}\{a_i,a_j\}=\{a_i,a_j\}&=&0,\nonumber\\
\{a_i,a_j^*\}&=&-\i\delta_{ij}.\label{pio5}\end{eqnarray}
Setting
  \beq\phi(w,w'):=\sum_i  a_i^*w_{i}+\sum_i  a_iw_{i}',\label{pio2}\eeq
we can rewrite (\ref{pio5}) as
\beq\left\{\phi(w_1,w_1')^*,\phi(w_2,w_2')\right\}=
-\i(w_1|w_2)+\i(w_1'|w_2')=-\i\big((w_1,w_1')|S(w_2,w_2')\big).\label{pio3}\eeq
In particular, $\phi(w,\bar w)$ are real, and
(\ref{pio3}) can be rewritten as
\beq\left\{\phi(w_1,\bar w_1)^*,\phi(w_2,\bar w_2)\right\}=2\Im(w_1|w_2)=
\Im\big((w_1,\bar w_1)|S(w_2,\bar w_2)\big).\label{pio3.}\eeq
Thus $S$ determines a symplectic structure on $\rr^{2m}$ (and
sometimes $S$ itself is called, incorrectly, a symplectic form).

\subsection{Symplectic transformations}
\label{Symplectic transformations}

In this subsection we recall some basic facts concerning the
symplectic and metaplectic group. We follow mostly \cite{DG}.

We say that an operator $R$ on $\cc^m\oplus\cc^m$
is  {\em symplectic} if it is  $J$-real
and preserves $S$:
\beq R^*SR=S.\label{eqi}\eeq
 We denote by $Sp(\rr^{2m})$
the group of all symplectic 
transformations.

Note that if $R$ is symplectic, then so is $R^*$. In fact, $\i S$ is
symplectic, and
\beq R^*=\i S R^{-1}(\i S)^{-1}.\eeq

The operator
 \beq
R=\left[\begin{array}{cc}p&q\\\bar q&\bar p\end{array}\right].
\label{written}\eeq
satisfies
 (\ref{eqi}) iff
\[
p^*p-q^\t \bar q=\one,\ \ p^*q-q^\t \bar p=0,
\]
\[
 pp^*-qq^*=\one,\ \ pq^\t -qp^\t =0.
\]

Note that 
\[
pp^*\geq \one,\ \ \ \ p^*p\geq \one.
\]
Hence $p^{-1}$ is well defined, and we can set
\begin{eqnarray}\label{dfd1}
d_1&:= &q^\t (p^\t )^{-1},\\\label{dfd2}
d_2&:=&q\bar p^{-1}.
\end{eqnarray}
We have $d_1^\t=d_1,$ $d_2=
d_2^\t$.

\subsection{Metaplectic transformations}

Let $U$ be a unitary operator
on $\Gamma_\s(\cc^m)$. Let $R$ be a symplectic transformation written
as (\ref{written}). We say that $U$ {\em  implements} $R$ if
\begin{eqnarray*}
U\hat a_i^*U^*&=&\hat a_j^*p_{ji}+\hat a_j\bar q_{ji},\\
U\hat a_iU^*&=&\hat a_j^*q_{ji}+\hat a_j\bar p_{ji}.
\end{eqnarray*}
$U$ will be called a {\em (Bogoliubov) implementer of $R$}.
Every symplectic transformation has an implementer, unique up to a
 a phase factor.
One can distinguish some canonical choices: the {\em natural implementer}
$U_R^\nat$, and a pair of {\em metaplectic implementers} 
$\pm U_R^\met$:
\begin{eqnarray}
U_R^{\nat}&:=&|\det
pp^*|^{-\frac{1}{4}}\e^{-\frac12\hat
  a^*(d_2)}\Gamma\bigl((p^*)^{-1}\bigr)\e^{\frac12\hat a(
  d_1)},\label{metap1}\\
\pm U_R^\met&:=&
\pm (\det
p^*)^{-\frac{1}{2}}\e^{-\frac12\hat
  a^*(d_2)}\Gamma\bigl((p^*)^{-1}\bigr)\e^{\frac12\hat a( d_1)},
\label{metap2}
\end{eqnarray}
see eg. Thm 11.33 and Def. 11.36 of \cite{DG}.

It is easy to see that the set of Bogoliubov implementers is a group.
It is  sometimes called the {\em  $c$-metaplectic group} $Mp^c(\rr^{2m})$.

It is a little less obvious, but also true, that the set of metaplectic Bogoliubov implementers is a subgroup
of   $Mp^c(\rr^{2m})$. It is called the {\em metaplectic group} $Mp(\rr^{2m})$.

We have a homomorphism $Mp^c(\rr^{2m})\ni U\mapsto R\in Sp(\rr^{2m})$, where to
$U$ implements $R$.

Various homomorphism related to the metaplectic group
can be described by the following diagram:
\beq
\begin{array}{ccccccccc}
&&1&&1&&1&&\\
&&\downarrow&&\downarrow&&\downarrow&&\\
1&\to&\zz_2&\to&U(1)&\to&U(1)&\to&1\\
&&\downarrow&&\downarrow&&\downarrow&&\\
1&\to&Mp(\rr^{2m})&\to&Mp^{\rm c}(\rr^{2m})&\to&U(1)&\to&1\\
&&\downarrow&&\downarrow&&\downarrow&&\\
1&\to&Sp(\rr^{2m})&\to&Sp(\rr^{2m})&\to&1&&\\
&&\downarrow&&\downarrow&&&&\\
&&1&&1&&&&
\end{array}
\eeq

\subsection{Positive symplectic transformations}

 {\em Positive symplectic transformation} are especially important. They satisfy
\beq
p=p^*,\ \ p>0,\ \ q=q^\t.\eeq
For positive transformations, $d_1$ equals $d_2$, and it  will be simply denoted by $d$. We have
\begin{eqnarray*}\label{dfd1a}
d&:= &q (p^\t )^{-1}.
\end{eqnarray*}
The natural implementer coincides in this case with one of the metaplectic
implementers:
\begin{eqnarray*}
U_R^{\nat}&:=&(\det
p)^{-\frac{1}{2}}\e^{-\frac12a^*(d)}\Gamma\bigl(p^{-1}\bigr)\e^{\frac12a(
  d)}.
\end{eqnarray*}

Positive symplectic transformations have special properties. In
particular, one can diagonalize them in an explicit way. We will need
this later on.

\bep Assume  that $R$ is positive symplectic and $\Ker(p-\one)=\{0\}$.
Then $q$ is invertible, so that we can define $u:=q|q|^{-1}$ with
$ |q|:=\sqrt{q^*q}$. Besides,
\beq M:=\frac{1}{\sqrt2}
\left[\begin{array}{cc}\one&-u\\u^*&\one\end{array}\right]
\label{propo}\eeq
is unitary
and diagonalizes $R$:
\beq
R=M\left[\begin{array}{cc}p+\sqrt{p^2-\one}
    &0\\0&\bar p-\sqrt{\bar p^2-\one}
  \end{array}\right]M^*.\label{dii}\eeq
\eep

\proof
We have the polar decomposition  $q=u|q|$.
$u$ is a unitary operator and we have $|\bar q|=u|q|u^*$

Now (\ref{dii}) follows using $u\bar pu^*=p$,
$|q|=\sqrt{\bar p^2-\one}$, $\sqrt{\one+|q|^2}=\bar p$.
\qed

\subsection{Classical quadratic Hamiltonians}
\label{Classical quadratic Hamiltonians}

It is easy to analyze generators of 1-parameter symplectic groups. In
fact,
 $\e^{\i tB}\in Sp(\rr^{2m})$ for any $t\in\rr$ iff
$BS$ is $J$-real and self-adjoint. All such operators can be written as
\beq
B=\left[\begin{array}{cc}h&-g\\\bar g&-\bar h\end{array}\right].
\label{gene}\eeq
where $h,g$ are $m\times m$ matrices satisfying  $h=h^*$, $g=g^\#$.
Note that $\i B$ is $J$-real, and
\beq SB=B^*S.\label{infi}\eeq

With every such an operator $B$ we associate another operator $A_B$ by
\beq\label{infi5}
A_B:=BS=\left[\begin{array}{cc}h&g\\\bar g&\bar h\end{array}\right],\eeq
As we noted above, $A_B$ is self-adjoint and $J$-real.
The corresponding {\em classical quadratic Hamiltonian} is the expression
\beq
H_B=\sum h_{ij} a_i^*a_j+\12\sum g_{ij}a_i^*a_j^*+\12\sum \bar
g_{ij}a_ia_j,\label{classi}\eeq
which can be viewed as a quadratic function on the classical phase space.
Moreover,
\beq\{H_B,\phi(w,w')\}=-\i\phi(w_1,w_1'),\ \ \ \ \ \left[\begin{array}{c}w_1\\w_1' 
  \end{array}\right]=B\left[\begin{array}{c}w\\w'\end{array}\right].\eeq 

Clearly, for any symplectic $R$,
\beq
A_{RBR^{-1}}=RA_BR^*.\eeq

In what follows  we
will  often abuse the terminology:  $A_B$ will also be called a {\em classical
   Hamiltonian} just as $H_B$. $B$ will be called a {\em symplectic
  generator}. Besides, we  will 
 often drop the subscript $B$ from $H_B$ and $A_B$.

\subsection{Quantum quadratic Hamiltonians}

Let $B$ be a symplectic generator of the form (\ref{gene}).

By a {\em quantization of $H_B$} (\ref{classi}) we will mean an operator on
$\Gamma_\s(\cc^m)$
of the form
\begin{eqnarray}
\hat H_B^c&:=&\sum h_{ij} \hat a_i^*\hat a_j+\12\sum g_{ij}\hat a_i^*\hat a_j^*+\12\sum \bar
g_{ij}\hat a_i\hat a_j+  c,
\label{hamo}\end{eqnarray}where
$c$ is an arbitrary real constant.
By an abuse of terminology, we will usually say that (\ref{hamo}) is a
{\em
  quantization of} $B$ (\ref{infi5}).
We will often drop the subscript $B$ from $\hat H_B^c$, and $c$ will be
replaced by other superscripts corresponding to some special choices.

Two  quantizations of $B$ are especially useful: the {\em Weyl} (or
{\em symmetric}) {\em quantization}
$\hat H_B^\w$ and
the {\em normally ordered} (or {\em Wick}) {\em quantization} $\hat H_B^\n$:
\begin{eqnarray*}
\hat H_B^\w&:=&\12\sum h_{ij} \hat a_i^*\hat a_j
+\12\sum h_{ij} \hat a_j\hat a_i^*+
\12\sum g_{ij}\hat a_i^*\hat a_j^*+\12\sum \bar g_{ij}\hat a_i\hat a_j,\\
\hat H_B^\n&:=&\ \ \ \ \sum h_{ij} \hat a_i^*\hat a_j\ \ \ \ +\12\sum g_{ij}\hat a_i^*\hat a_j^*+\12\sum \bar
g_{ij}\hat a_i\hat a_j.
\end{eqnarray*}
Here is the relation between these two quantizations:
\beq \hat H_B^\w=\hat H_B^\n+\12\Tr h.\label{fs0}\eeq

Note a special relationship of the Weyl quantization to the
metaplectic group (defined in Subsect. \ref{Symplectic transformations}):
for any $B$, $\e^{\i t\hat H_B^\w}$ belongs to $Mp(\rr^{2m})$, see
eg. Thm 11.34 of \cite{DG}.
Besides, if $R$ is symplectic and $U_R$ is its implementer, then
\beq U_R\hat H_B^\w U_R^*=\hat H_{RBR^{-1}}^\w.\label{metap}
\eeq

\subsection{Diagonalization of quadratic Hamiltonians}

In this subsection we show that if $A_B>0$, then  $A_B$ can be {\em diagonalized}. 
By this we mean that we can find a symplectic transformation
$R$ that kills  off-diagonal terms of $A_B$:
\beq A_B=R\left[\begin{array}{cc}
h_\dg&0\\0& \bar h_\dg\end{array}\right]R^*
\label{infi9a}\eeq
for some $h_\dg$.
Of course, $h_\dg$ has to be positive.

Clearly, this is equivalent to {\em diagonalizing} $B$, that is to killing its off-diagonal
terms: of  $B$, that is,
\beq B=R\left[\begin{array}{cc}
h_\dg&0\\0& -\bar h_\dg\end{array}\right]R^{-1}.
\label{infi9b}\eeq

On the quantum level, this is equivalent to finding a
unitary operator $U$  that removes
double annihilators and double creators. Then
the free constant equals the infimum
 of the quantum Hamiltonian:
\begin{eqnarray*}
U^*\hat H^\w U&=&\d\Gamma(h_\dg)+E^\w,\\
U^*\hat H^\n U&=& \d\Gamma(h_\dg)+E^\n.\end{eqnarray*}

As a preparation for a construction of a diagonalizing operator, let
us prove the following proposition. In this proposition we will
use the function $\sgn{t}:=\begin{cases}1&t>0,\\
0&t=0,\\
-1&t<0.
\end{cases}$

\begin{proposition} Suppose that $A_B>0$.
  \ben\item  The operator $B$ has only real nonzero eigenvalues. Therefore, $\sgn$
can be interpreted as a holomorphic function on a neighborhood of $\sp
B$, and
we can define
$\sgn(B)$ by the standard holomorphic functional calculus.\item
A symplectic transformation $R$ diagonalizes $B$ iff
  \beq  \sgn(B)=RSR^{-1}.\label{pow}\eeq
  \een
\label{prio}  \eep

\proof
It is useful to endow the space $\cc^m\oplus\cc^m$ with the scalar
product given by the positive operator $SAS$.
More precisely, if $v=(v_1,v_2), w=(w_1,w_2)\in \cc^m\oplus\cc^m$, we set
\beq(v|w)_\en=(v|SASw)=
(v_1|hw_1)-(v_1|gw_2)-(v_2|\bar
gw_1)+(v_2|\bar h w_2).\label{ene}\eeq
(\ref{ene}) is sometimes called the {\em
  energy scalar product}.

Note that we also have the original scalar product
\[(v|w)=(v_1|w_1)+(v_2|w_2),\]
which is used for basic
notation such as the Hermitian adjoints.

First note that $B$ is self-adjoint in the energy scalar product and
has a zero nullspace. Indeed
\begin{eqnarray*}
  (v|Bw)_\en&=&(v|SASASw)\\
  =\ (ASv|SASw)&=&(Bv|w)_\en,\\
  (Bv|Bv)_\en&=&(v|SASASASv)>0,\ \ v\neq0.
  \end{eqnarray*}
This shows (1).

Now let $R$ be symplectic. Set
\[ B_\dg:= R^{-1}BR,\ \ \ A_\dg:=B_\dg S= R^{-1}AR^{*-1}.\]
Then, by functional calculus,
\beq\sgn(B)=R\sgn( B_\dg)R^{-1}.\label{pqi}\eeq

 $R$ diagonalizes $A$ iff
\beq A_\dg
=
\left[\begin{array}{cc}
h_\dg&0\\0& \bar h_\dg\end{array}\right],\ \ \  B_\dg=\left[\begin{array}{cc}
h_\dg&0\\0& -\bar h_\dg\end{array}\right].\label{war}\eeq
 $A$ is strictly positive, hence so are $ A_\dg$ and
$h_\dg$. Therefore,
\beq
\sgn(B_\dg)=S.\label{pow1}\eeq
Together with (\ref{pqi}), this implies
(\ref{pow}).

Conversely, suppose that (\ref{pow}) holds.
Together with (\ref{pqi}), this implies (\ref{pow1}). Hence  $B_\dg$ is
diagonal.
\qed

It is possible to find a distinguished positive symplectic transformation $R$
diagonalizing $B$. 

\bet Suppose that $A_B>0$. \ben
\item $\i\,\sgn(B)$ is symplectic.
\item
  $R_0:=\sgn( B)S$  is symplectic and has positive eigenvalues.
 \item Using holomorphic calculus and the
principal  square root (which for positive arguments has
positive values),
 define
\beq
R:=R_0^{\frac12}.\label{war1}\eeq
Then
$R$ is positive, symplectic and diagonalizes $B$.
\item Here is an alternative formula for $R_0$, where the square root
  can be interpreted in terms of functional calculus for self-adjoint
  operators: 
\beq R_0=SA_B^{-\frac12}\big(A_B^{\frac12}SA_BSA_B^{\frac12}\big)^{\frac12}A_B^{-\frac12}S.\label{infi3}\eeq
  \een
\label{th1}\eet

\proof $B$ satisfies 
(\ref{infi}). Hence for any   function $f$ holomorphic on the spectrum of $B$
\beq
Sf(B)S^{-1}=f(B^{*}).\label{infi1}\eeq
In particular
\begin{equation}
  S\sgn(B)S^{-1}=\sgn(B^*).\label{inffi7}\eeq
But $\sgn$ is real, hence $\sgn(B^*)=\sgn(B)^*$.
Besides, away from $0$ we have $\sgn(t)=\sgn(t)^{-1}$.
Hence, $\sgn(B)=\sgn(B)^{-1}$.
Therefore, (\ref{inffi7}) can be rewritten as
\begin{equation}
  S\sgn(B)S^{-1}=\sgn(B)^{*-1}.\label{inffi8}\eeq
  Hence,
\beq\big(\i\,\sgn(B)\big)^*S\i\,\sgn(B)=S.\eeq
This means that  $\i\,\sgn(B)$ preserves $S$.
Besides, \beq
  \i\,\sgn
  B=\big(-(\i B)^2\big)^{\frac12}(\i B)^{-1}\eeq
  is also $J$-real.
  Thus we have shown that $\i\,\sgn(B)$ is  symplectic.

  $-\i S$ is also symplectic. 
Therefore, so is $R_0=\big(\i\,\sgn(B)\big)\big(-\i S\big)$.

Now, 
\begin{eqnarray}  R_0&=&(B^2)^{\frac12}B^{-1}S\\
  &=&(ASAS)^{\frac12}SA^{-1}S\\
  &=&SA^{-\frac12}\big(A^{\frac12}SASA^{\frac12}\big)^{\frac12}A^{-\frac12}S.
\end{eqnarray} 
Therefore, (\ref{infi3}) is true and
$R_0$ is a positive self-adjoint operator for  the original
scalar product. Hence it has positive eigenvalues.

 $R_0=R_0^*$ and  $R_0$ is symplectic. Hence,
\[SR_0S^{-1}=R_0^{-1}.\]
Hence for any Borel function $f$,
\[Sf(R_0)S^{-1}=f(R_0^{-1}).\]
Choosing $f$ to be  the (positive) square root we obtain
\[SR_0^{\frac12}S^{-1}=R_0^{-\frac12}.\]
Thus  $R:=R_0^{\frac12}$ is symplectic, positive  and self-adjoint for
the original scalar product.

Now
\[\sgn(B)=R^2S=RSR^{-1}.\]
Hence (\ref{pow}) is true.
\qed


\subsection{Positive  Weyl Bogoliubov  Hamiltonians}

\bet\ben\item If $A_B\geq0$, then the Weyl quantization of $B$
is positive. Hence
all quantizations of $B$ are bounded
from below.\item If $B$ possesses a quantization that is bounded
from below, then $A_B\geq0$.
\een\label{pok1}
\eet

\proof
(1) Let $A\geq0$. Then there exists a symplectic
transformation $R$ and a decomposition
$\cc^m=\cc^{m_1}\oplus\cc^{m-m_1}$ such that $RAR^*$
decomposes into the direct sum of the following two terms:
\begin{eqnarray*}
&\left[\begin{array}{cc}h_\dg&0\\0&\bar h_\dg\end{array}\right]&\hbox{
  on }\cc^{m_1}\oplus\cc^{m_1},\\
&\frac12\left[\begin{array}{cc}\one&\one\\\one&\one\end{array}\right]&\hbox{
  on }\cc^{m-m_1}\oplus\cc^{m-m_1},
\end{eqnarray*}
where $h_\dg\geq0$ and can be assumed to be
diagonal. This is a well-known fact, proven eg. in
\cite{Ho1,DG}. It is a very special  case of a more general and more
complicated classification of quadratic forms on a symplectic space called
Williamson's Theorem, proven eg. in \cite{Wi,Ho}. If we strengthen the
assumption and demand that $h>0$, it follows also from the
diagonalizability of $A$ (Theorem \ref{th1}).
Thus, after an application of the transformation $R$, and a
diagonalization of $h_\dg$, the classical
Hamiltonian becomes
\beq
H_{RBR^{-1}}=\sum_{i=1}^{m_1}h_{\dg,ii} a_i^*a_i+\sum_{m_1+1}^{m}\frac12(
a_i^*+a_i)^2 .\eeq
After application of an implementer of $R$,  the quantum Weyl Hamiltonian
becomes
\beq
U_R\hat H_B^\w U_R^*=
\hat H_{RBR^{-1}}^\w=\sum_{i=1}^{m_1}\frac12h_{\dg,ii}(\hat a_i^*\hat a_i+\hat a_i\hat a_i^*)
+\sum_{m_1+1}^{m}\frac12(
\hat a_i^*+\hat a_i)^2 .\label{pok}\eeq
Thus $\hat H_B^\w$ is  positive.

(2)
Consider the family of coherent vectors
\beq\Omega_w:=\e^{\hat a^*(w)-\hat a(w)}\Omega,\ \ \ w\in\cc^m.\eeq
Note that
\beq \e^{-\hat a^*(w)+\hat a(w)}\hat a_i^*\e^{\hat a^*(w)-\hat a(w)}=\hat a_i^*+\bar w_i,\ \ \
 \e^{-\hat a^*(w)+\hat a(w)}\hat a_i\e^{\hat a^*(w)-\hat a(w)}=\hat a_i+ w_i.\label{pom}\eeq

Obviously, if one of quantizations of $B$ is bounded from below, then
so are all of them.
Let $\hat H_B^\n$ be bounded from below by $-c$. Then, using (\ref{pom}), we obtain
\begin{eqnarray}
  -c&\leq&(\Omega_w|\hat H_B^\n\Omega_w)\\
  &=&\Big(\Omega|\e^{-\hat a^*(w)+\hat a(w)}\hat
  H_B^\n\e^{\hat a^*(w)-\hat a(w)}\Omega\Big)\\
  &=&\sum h_{ij}\bar w_iw_j+\frac12\sum g_{ij}\bar w_i\bar w_j+
  \frac12\sum \bar g_{ij} w_i w_j.
\end{eqnarray}
Thus the classical Hamiltonian is quadratic
polynomial and is
 bounded from below. But if a quadratic polynomial is bounded from below, then
it is nonegative.
\qed

Note that by the above theorem,  every $B$ satisfying  $A_B\geq0$, beside $\hat H_B^\w$ and
$\hat H_B^\n$, possesses another natural
quantization: the {\em zero infimum quantization} $\hat H_B^\z$ fixed by the
condition
\beq\inf \hat H_B^\z=0.\eeq

The infimum of the Weyl Bogoliubov Hamiltonians can be computed from
several formulas described in the following theorem borrowed from
 \cite{BD,DG}:

\bet Assume that $A_B\geq0$. Then
\begin{eqnarray}E_B^\w:=\inf \hat H_B^\w&
=&\frac{1}{4}\Tr
\sqrt{B^2}\label{fun1}
\\
&=&\frac14\Tr\left[\begin{array}{cc}
h^2-g\bar g& -hg+g \bar h\\ \bar g h- \bar h \bar g&\bar h^{2}-\bar gg\end{array}\right]
^{\frac12}\label{fun2}
\\
&=& \frac{1}{4}\Tr
\sqrt{A^{\frac12}SASA^{\frac12}}
\label{funct}
\\
&
=&\frac14\Tr\int\frac{B^2}{B^2+\tau^2}\frac{\d\tau}{2\pi}.\label{func3}
\end{eqnarray}\label{func1}
\eet

\proof  Let $R$ be as in the proof of Theorem \ref{pok1}. Clearly,
\begin{eqnarray}\inf(\hat a_i^*\hat a_i+\hat a_i\hat a_i^*)&=&1,\\
  \inf(\hat a_i^*+\hat a_i)^2&=&0.
\end{eqnarray}  
Hence, by (\ref{pok}),
\begin{eqnarray}
\inf\hat H_B^\w=\inf U_R\hat H_B^\w U_R^*=\inf\hat H_{RBR^{-1}}^\w&=&\frac12\sum
h_{\dg,ii}=\frac12\Tr h_\dg\\&=&
\frac14 \Tr\sqrt{ B_\dg^2}=
\frac14 \Tr R\sqrt{ B^2}R^{-1}=
\frac14 \Tr\sqrt{ B^2}
.\end{eqnarray}
 This gives (\ref{fun1}), which implies
(\ref{fun2}) and (\ref{funct}). 

(\ref{func3}) follows by an application of the identity
(\ref{iden1}).
\qed


\subsection{Infimum of normally ordered Hamiltonians}
\label{Infimum of normally ordered Hamiltonians}

 Here are a few formulas
 for the infimum of the normally ordered Hamiltonian:

 \bet Assume that $A_B\geq0$. Then
\begin{eqnarray}
  E_B^\n:=\inf\hat H_B^\n&=&E_B^\w-\12\Tr h\label{fs1}\\
&=&\frac14\Tr\big(\sqrt{B^2}-\sqrt{B_0^2}\big)\label{fs2}\\
  &=&\frac14
  \Tr\Biggl(\left[\begin{array}{cc}
h^2-g\bar g& -hg+g \bar h\\ \bar g h- \bar h \bar g&\bar h^{2}-\bar gg\end{array}\right]
^{\frac12}
-\left[\begin{array}{cc}
    h&0\\0& \bar h\end{array}\right]\Biggr)\label{fs3}\\
&=&\frac18\int_0^1\d\sigma
\Tr\frac{B_\sigma}{\sqrt{B_\sigma^2}}G
.\label{fs4}\\
&=&\frac18\int_0^1\d\sigma
\Tr A_\sigma^{\frac12}(A_\sigma^{\frac12}SA_\sigma
SA_\sigma^{\frac12})^{-\frac12}
 A_\sigma^{\frac12}GS\label{fs4a}\\
&=&\frac18\int_0^1\d\sigma\int\frac{\d\tau}{2\pi}(1-\sigma)
\Tr
\frac{1}{(A_\sigma+\i\tau S)}SG\frac{1}{(A_\sigma+\i\tau S)}SG.\label{fs5}
\end{eqnarray}
where
\begin{eqnarray}
G&:=&B-B_0=\left[\begin{array}{cc}0&-g\\\bar g&0
  \end{array}\right],\label{fa2}\\
A_\sigma&=&A_0+\sigma GS=
\left[\begin{array}{cc}h&\sigma
      g\\\sigma \bar g&\bar
      h\end{array}\right],\label{fa1}\\
B_\sigma&:=&B_0+\sigma G=
\left[\begin{array}{cc}h&-\sigma
      g\\\sigma \bar g&-\bar
      h\end{array}\right],\ \ \ \sigma\in\rr.\label{fa3}
\end{eqnarray}
\label{thm-nor}
\eet

\proof
(\ref{fs1}), (\ref{fs2}) and (\ref{fs3}) follow immediately from
Theorem \ref{func1} and (\ref{fs0}).

Starting  from (\ref{fs2}), let us prove (\ref{fs4}), which
immediately implies (\ref{fs4a}):
\begin{eqnarray}
&&\frac14\Tr\big(\sqrt{B^2}-\sqrt{B_0^2}\big)\\
  &=&\frac14\int\Tr\Big(\frac{B^2}{B^2+\tau^2}-\frac{B_0^2}{B_0^2+\tau^2}\Big)\frac{\d\tau}{2\pi}\\
  &=&-\frac14\int\Tr\Big(\frac{1}{B^2+\tau^2}-\frac{1}{B_0^2+\tau^2}\Big)
  \frac{\tau^2\d\tau}{2\pi}\\
  &=&  -\frac14\int_0^1\d\sigma\int\frac{\d}{\d\sigma}
  \Tr\frac{1}{B_\sigma^2+\tau^2}
  \frac{\tau^2\d\tau}{2\pi}\\
  &=&  \frac14\int_0^1\d\sigma\int
  \Tr\frac{1}{B_\sigma^2+\tau^2}(B_\sigma G+GB_\sigma)\frac{1}{B_\sigma^2+\tau^2}
  \frac{\tau^2\d\tau}{2\pi}\\
  &=&  \frac12\int_0^1\d\sigma\int
  \Tr\frac{B_\sigma}{(B_\sigma^2+\tau^2)^2}G
  \frac{\tau^2\d\tau}{2\pi}\\
  &=&\frac18\int_0^1\d\sigma
\Tr\frac{B_\sigma}{\sqrt{B_\sigma^2}}G,
    \end{eqnarray}
where at the end we used the identity (\ref{app2}).

Now,  starting from (\ref{fs4}),
we prove (\ref{fs5}):
\begin{eqnarray}
 && \frac18\int_0^1\d\sigma\int\frac{\d\tau}{2\pi}
  \Tr\frac{B_\sigma}{B_\sigma^2+\tau^2}G\\
  &=& \frac1{16}\int_0^1\d\sigma\int\frac{\d\tau}{2\pi}
  \Tr\Big(\frac{1}{(B_\sigma+\i\tau)}+\frac{1}{(B_\sigma-\i\tau)}\Big)
  G\\
  &=& \frac18\int_0^1\d\sigma\int\frac{\d\tau}{2\pi}
  \Tr\frac{1}{(B_\sigma+\i\tau)}
  G\\
   &=& \frac18\int_0^1\d\sigma\int_0^\sigma\d\sigma_1\int\frac{\d\tau}{2\pi}
  \frac{\d}{\d\sigma_1}\Tr\frac{1}{(B_{\sigma_1}+\i\tau)}
  G\label{ppp3}\\
  &=& -\frac18\int_0^1(1-\sigma)\d\sigma\int\frac{\d\tau}{2\pi}
  \Tr\frac{1}{(B_\sigma+\i\tau)}G\frac{1}{(B_\sigma+\i\tau)}G\label{ppp2}
  \\
  &=& -\frac18\int_0^1(1-\sigma)\d\sigma\int\frac{\d\tau}{2\pi}
  \Tr\frac{1}{(A_\sigma+\i\tau S)}SG\frac{1}{(A_\sigma+\i\tau S)}SG\label{ppp1}
.
\end{eqnarray}

In (\ref{ppp3})$\Rightarrow$(\ref{ppp2}) we used
\beq \frac{\d}{\d\sigma_1}
\frac{1}{(B_{\sigma_1}+\i\tau)}=-\frac{1}{(B_{\sigma_1}+\i\tau)}G\frac{1}{(B_{\sigma_1}
  +\i\tau)}.\eeq
\qed

\subsection{Loop expansion}
\label{Loop expansion}
Suppose now that 
\beq
B_0=\left[\begin{array}{cc}h_0&0\\0&-\bar
      h_0\end{array}\right],\label{fa1a}
\eeq
is a ``free'' symplectic generator. We assume that $h_0>0$.
Note that we allow $h_0$ to be different from $h$.

We set
\begin{eqnarray}
  A_0:=B_0S&=&\left[\begin{array}{cc}h_0&0\\0&\bar
      h_0\end{array}\right],\\ 
V:=B^2-B_0^2&=&\left[\begin{array}{cc}h^2-h_0^2-g\bar g&-hg+g\bar
    h\\\bar gh-\bar h
  \bar  g&\bar h^2-\bar h_0^2-\bar gg
  \end{array}\right].\label{fa2a}
\end{eqnarray}

(\ref{func3}) can be rewritten as
 \begin{eqnarray*}E^\w&
=&\frac14\Tr\int\frac{B_0^2}{B_0^2+\tau^2}\frac{\d\tau}{2\pi}
+\frac14\Tr\int\frac{1}{B^2+\tau^2}V\frac{1}{B_0^2+\tau^2}\tau^2\frac{\d\tau}{2\pi}\\
&=&\sum_{j=0}^kL_j+
\frac14\Tr\int\frac{(-1)^{k}}
            {B_0^2+\tau^2}V\frac{1}{B^2+\tau^2}\Big(V\frac{1}{B_0^2+\tau^2}\Big)^k
            \tau^2\frac{\d\tau}{2\pi}\\
&=&\sum_{j=0}^\infty L_j,
\end{eqnarray*}
 where 
\begin{eqnarray}L_0&=&\frac14\Tr\int\frac{B_0^2}{B_0^2+\tau^2}\frac{\d\tau}{2\pi}
=\frac14\Tr|B_0|=\frac12\Tr h_0,\\
L_j&=&\frac14\Tr\int\frac{(-1)^{j+1}}{B_0^2+\tau^2}\Big(V\frac{1}{B_0^2+\tau^2}\Big)^j
\tau^2\frac{\d\tau}{2\pi}\\
&=&\frac14\Tr\int\frac{(-1)^{j+1}}{2j}\Big(V\frac{1}{B_0^2+\tau^2}\Big)^j
\frac{\d\tau}{2\pi},\ \ \ \ j=1,2,\dots.\label{cyclic}
\end{eqnarray}
The last identity for $L_j$ follows by a cyclic relocation of
operators under the trace and by an application of integration
by parts.

We can further simplify the formula for $L_1$:
\beq
L_1=\Tr\int\frac{1}{8}V\frac{1}{B_0^2+\tau^2}
\frac{\d\tau}{2\pi}
=\frac18\Tr V\frac{1}{|B_0|}=\frac14\Tr(h^2-h_0^2-g\bar
g)h_0^{-1}.\label{simpli}\eeq

The constant $L_j$ arises in the diagramatic expasion as the evaluation
of the loop with $j$ vertices.
To see this, introduce the ``time variable'' $t$ and the ``Feynman propagator''
\[G(t):=\frac{\e^{-|B_0|t}}{2|B_0|}.\]
Clearly, $\tau$ can be interpreted as the ``energy variable'' and
\[\frac{1}{B_0^2+\tau^2}=\int G(t)\e^{\i t\tau}\d t.\]
Therefore,
\begin{align}
  &L_j\ =
 \ \frac14\int\d t_{j-1}\cdots\int\d t_1\Tr VG(t_j-t_1)VG(t_1-t_2)\cdots
  VG(t_{j-1}-t_j)\\ 
&=\lim_{T\to\infty}\frac{1}{2T}
  \frac14\int_{-T}^T\d t_{j}\int_{-T}^T\d t_{j-1}\cdots\int_{-T}^T
  \d t_1\Tr VG(t_j-t_1)VG(t_1-t_2)\cdots   VG(t_{j-1}-t_j).
\end{align}

\subsection{Renormalization I}
\label{Renormalization I}

Note that in 
general $V$ (\ref{fa2a}) contains terms of the 1st and 2nd
order. Explicitly, let $\lambda$ be a ``coupling constant''. Let
$h=h_0+\lambda h_1$ and replace $g$ with $\lambda g$ (to keep track of
the order of perturbation). Then $V=\lambda V_1+\lambda^2V_2$, where
\begin{eqnarray}
V_1&:=&\left[\begin{array}{cc}h_0h_1+h_1h_0&-h_0g+g\bar
    h_0\\\bar gh_0-\bar h_0
  \bar  g&\bar h_0\bar h_1+\bar h_0\bar h_1
  \end{array}\right],\label{fa2aa}\\
V_2&:=&\left[\begin{array}{cc}h_1^2-gg^*&-h_1g+g\bar h_1
    \\\bar gh_1-\bar h_1\bar g&\bar h_1^2-g^*g
  \end{array}\right].\label{fa2a--}
\end{eqnarray} 

 We can expand $E^\w$ wrt the coupling constant $\lambda$:
 \beq E^\w=\sum_{n=0}^\infty \lambda^nE_n.\label{expa1}\eeq
 We have,
 \beq L_0=E_0=\frac12\Tr h_0.\eeq
However, in general, $L_n$ of higher orders differ from
$\lambda^nE_n$.

There are situations when it is natural to introduce
the {\em renormalized vacuum energy}
\begin{eqnarray}
  E^\ren&:=E^\w-E_0-\lambda E_1-\lambda^2E_2=
  \sum\limits_{n=3}^\infty\lambda^n E_n.\label{expa2}\end{eqnarray}
and  the {\em renormalized Hamiltonian}
\beq  \hat H^\ren:=\hat H^\w-E_0-\lambda E_1-\lambda^2E_2,\eeq
so that $E^\ren=\inf  \hat H^\ren$.
The numbers $E_0$, 
$E_1$ and $E_2$ can be called {\em counterterms}.

The above constuctions are natural  e.g. in the theory of charged scalar fields in external electromagnetic potentials. In this case, $E_0,E_1, E_2$ are infinite. $\hat H^\ren$ is usually also ill defined. However $E^\ren$ is typically finite. 
 Thus we have a somewhat paradoxical situation: the Hamiltonian does not exists, however the  ``infimum of the Hamiltonian'' is well defined.

\subsection{Renormalization II}
\label{Renormalization II}

Suppose now that \beq h_1^2=g\bar g,\ \ \ \ h_1g=g\bar h_1.\label{merits}\eeq
(\ref{merits})  implies $V_2=0$. Therefore, the
loop expansion coincides with 
the expansion into powers of $\lambda$. Putting $\lambda=1$, we thus
have
\beq E_n=L_n,\ \ \ n=0,1,\dots\eeq
We can compute the loop with one vertex:
\beq L_1=\frac14\Tr(h_0h_1+h_1h_0)h_0^{-1}=\frac12\Tr h_1.\eeq
Thus
\beq L_0+L_1=\frac12\Tr(h_0+h_1)=\frac12\Tr h.\eeq
Therefore,  the loop expansion for the infimum of the
normally ordered Hamiltonian amounts to omitting $L_0$ and $L_1$:
\beq\inf E^\n=E^\w-\frac12\Tr h=\sum_{n=2}^\infty L_n.\eeq

Note that $L_1$, and especially $ L_0$, are often infinite. Sometimes,
$L_2$ is infinite as well. Then we can renormalize the
vacuum energy even further, introducing
\begin{eqnarray}
  E^\ren&:=&E^\w-L_0-L_1-L_2=
  \sum\limits_{n=3}^\infty L_n\\
&=&-\frac14\int\Tr\frac{1}{B_0^2+\tau^2}V\frac{1}{B^2+\tau^2}\Big(V\frac{1}{B_0^2+\tau^2}\Big)^2
\tau^2\frac{\d\tau}{2\pi}\label{eren}
.\end{eqnarray}
We  can  also introduce the renormalized Hamiltonian
\beq  \hat H^\ren:=\hat H^\w-L_0-L_1-L_2,\eeq
so that
\beq E^\ren=\inf  \hat H^\ren.\label{info}\eeq

The situation described in this subsection is typical for a charged
particle in an external electrostatic potential (without a vector
potential), as well as for a neutral scalar particle with a masslike
perturbation \cite{De}. One can then often introduce 
the renormalized Hamiltonian
$ \hat H^\ren$, which is a well-defined self-adjoint operator so that
(\ref{info}) holds.

\section{Arbitrary dimensions, basis independent formalism}
\init

In this section we consider Bogoliubov Hamiltonians in any
dimension. Unlike in the previous section, we will use a basis
independent notation.

We will use the standard notation for the Hilbert-Schmidt and trace
class norm:
\beq\|g\|_2:=\sqrt{\Tr g^*g},\ \ \ \ \ \|g\|_1:=\Tr\sqrt{ g^*g}.\eeq

\subsection{Doubled space in abstract setting}

Let $\cW$ be a Hilbert space. $\cW$ will serve as the {\em 1-particle
  space}.

Let
$\bar\cW$ be another Hilbert space with a fixed
antiunitary map $\chi:\cW\to\bar\cW$.
$\bar\cW$ will be called the {\em complex conjugate
  of} $\cW$,

We will often use the doubled space $\cW\oplus\bar\cW$ equipped with the
conjugation
\beq
J=\left[\begin{array}{cc}0&\chi^{-1}\\\chi&0\end{array}\right].\eeq
A {\em $J$-real} operator is an operator on $\cW\oplus\bar\cW$ commuting with
$J$. Bounded $J$-real operators have the form
\beq
R=\left[\begin{array}{cc}p&q\\\chi q\chi&\chi p\chi^{-1}
  \end{array}\right],\eeq
for some $p\in B(\cW)$, $q\in B(\bar\cW,\cW)$.

$J$-real operators leave invariant the real subspace of vectors
\[\left[\begin{array}{c}w\\\chi w\end{array}\right],\ w\in\cW,\]  which we will
denote $\cY$.
Note also that every $J$-real operator in $ B(\cW\oplus\bar \cW)$
restricts to an operator $B(\cY)$, and conversely, each operator in 
$B(\cY)$ extends uniquely to an operator in $B(\cW\oplus\bar\cW)$.

In what follows, we will usually write $\bar w$ for $\chi w$. We will 
write $\bar p$, $\bar q$ for
$\chi p\chi^{-1}$ and $\chi q\chi$.
We will write $p^\#$, $q^\#$ for
$\chi p^*\chi^{-1}$ and $\chi^{-1} q^*\chi^{-1}$.
In Appendix \ref{Complex conjugate space}
we explain why it is natural to use this
simplified notation.

To reduce the formalism of this section to that
of  Section \ref{Finite dimensions, basis dependent
  formalism} it suffices  to set
$\cW=\cc^m$ and replace $\chi$ with the complex conjugation.

\subsection{Fock spaces}
\label{Fock spaces}

If $\cD$ is a vector space of any dimension (with or without a Hilbert
space structure), then we can introduce its {\em algebraic $n$th symmetric power},
denoted by $\otimesal_\s^n\cD$ and the
the {\em algebraic bosonic Fock space}
\[
\Gammal_\s(\cW):= \oplusal_{n=0}^{\infty} {\otimesal}^n_\s\cD,
\]
which is the space of finite
symmetric tensor products of vectors of $\cD$ \cite{DG}.
If $\cW$ is a Hilbert space, then we prefer to use the Hilbert space
versions of the above constructions. Thus 
 $\otimes_\s^n\cW$ will denote the  {\em $n$th symmetric tensor power of
$\cW$ in the sense of Hilbert spaces} and, as usual,
 the {\em bosonic Fock space over the one-particle space $\cW,$} is
 defined as
\[
\Gamma_\s(\cW):= \mathop{\oplus}\limits_{n=0}^{\infty} \otimes^n_\s\cW.
\]
$\Omega:= (1, 0,\cdots )$  denotes the vacuum vector and 
\begin{eqnarray*}
\Gamma^{{\rm fin}}_\s(\cW)&:=&\oplusal_{n=0}^{\infty} \otimes^n_\s\cW\\
&=&\big\{(\Psi^{0},\cdots,\Psi^{n},\cdots) \in\Gamma_\s(\cW) \ | \ \Psi^{n}=0 \ {\rm for \ all \ but \ a \ finite \ number \ of \ }n\big\}, 
\end{eqnarray*}
is the {\em finite particle bosonic Fock  space}.

Note that if $\cD$ is dense in $\cW$, then $\Gammal_\s(\cD)$ is
dense in  $\Gamma_\s(\cW)$.


If $h$ is an operator on $\cW,$ $\d\Gamma(h)$ will denote 
$$
\d\Gamma(h)\lceil_{\otimes^n_\s \cW}:= \sum_{j=1}^n
\underbrace{1\otimes \cdots \otimes 1}_{j-1}\otimes h\otimes
\underbrace{1\otimes \cdots \otimes 1}_{n-j}
\lceil_{\otimes^n_\s \cW}.
$$

If $q$ is an operator on $\cW$ of norm less than $1$,
 we define
$\Gamma(q) : \Gamma_\s(\cW) \to \Gamma_\s(\cW)$ by 
\[
\Gamma(q)\lceil_{\otimes^n_\s \cH} := q\otimes \cdots \otimes q
\lceil_{\otimes^n_\s \cH}. 
\]

\subsection{Quadratic forms on Fock spaces}

For any  operator $h$ on $\cW$ such that $h\geq c$,
its {\em form domain} is defined
  as
  \beq\Dom(|h|^{\frac12})=(\one+|h|)^{-\frac12}\cW.\eeq
  For $ w_1,w_2\in \Dom(|h|^{\frac12})$, we can define $(w_i|hw_2)$.
$\Dom(|h|^{\frac12})$ is a Hilbert space for the scalar product
  $\big(w_1|(h+c+\one)w_2\big)$.
  We say that $\cD$ is a {\em form core for $h$} if it is a dense
  subspace of the form domain of $h$.

  \bel Suppose that  $h\geq0$ and $\cD$ is a form core of $h$.
  Then  $\Gammal_\s(\cD)$ is a form 
core of
$\d\Gamma(h)$. \label{pqy}\eel

\proof
It is easy to see that
\beq\one+\d\Gamma(h)\leq\Gamma(\one+h).\eeq
Hence,
\beq\Gamma(\one+h)^{-\frac12}\Gamma_\s(\cW)
\subset\big(\one+\d\Gamma(h)\big)^{-\frac12}\Gamma_\s(\cW).\label{core1}
\eeq

Let $\Psi\in\big(\one+\d\Gamma(h)\big)^{-\frac12}\Gamma_\s(\cW).$ Set
\beq\Psi_n:=\one_{[0,n]}\big(\Gamma(\one+h)\big)\Psi.\eeq
By the spectral theorem and the fact that $\one+\d\Gamma(h)$ and
$\Gamma(\one+h)$ commute with one another,
$\Psi_n\in\Dom\Gamma(\one+h)^{\frac12}$ and $\Psi_n\to\Psi$ in
$\big(\one+\d\Gamma(h)\big)^{-\frac12}\Gamma_\s(\cW)$. Hence
\beq
\Gamma(\one+h)^{-\frac12}\Gamma_\s(\cW)\ \hbox{ is dense in }\ 
\big(\one+\d\Gamma(h)\big)^{-\frac12}\Gamma_\s(\cW).\label{dense1}\eeq

Now $\cD$ is dense in $(\one+h)^{-\frac12}\cW$. Hence
\beq
\Gammal_\s(\cD)\ \hbox{ is dense in }\ 
\Gamma_\s\big((\one+h)^{-\frac12}\cW\big)=
\Gamma(\one+h)^{-\frac12}\Gamma_\s(\cW).\label{dense2}\eeq

Putting together (\ref{dense1}) and (\ref{dense2}) we obtain
\beq
\Gammal_\s(\cD)\ \hbox{ is dense in
}\ \big(\one+\d\Gamma(h)\big)^{-\frac12}\Gamma_\s(\cW).
\label{dense3}\eeq 
But the RHS of (\ref{dense3}) is the form domain of 
$\d\Gamma(h)$.
\qed

\subsection{Creation/annihilation operators}

For any $w\in \cW,$ $\hat a(w)$ and $\hat a^*(w)$ denote the usual
annihilation/creation operators
 \begin{eqnarray}
\hat a^*(w)\Psi&:=&\sqrt{n+1}w\otimes_\s\Psi, \qquad
\Psi\in\otimes_\s^n\cW,\label{anni1}
\\
\hat a(w)\Psi&:=&\sqrt{n+1} ( w|\, \otimes1^{\otimes n}\,\Psi, \qquad \Psi\in\otimes_\s^{n+1}\cW,\label{anni}
\end{eqnarray}
These operators, originally well defined
on $\Gamma^{{\rm fin}}_\s(\cW)$, extend to closed operators on
on $\Gamma_\s(\cW)$. We set
  \beq\hat\phi(w,\bar w'):=\hat a^*(w)+\hat a(w').\label{pio2a}\eeq
Note that $\hat\phi(w,\bar w)$ are self-adjoint. One can also introduce
the so-called {\em Weyl 
operators}  $\e^{\i\hat\phi(w,\bar w)}$.

\ber Sometimes we may want to define creation/annihilation operators for
$w$ that do not belong to $\cW$, but are functionals, possibly unbounded, with
domain $\cD\subset \cW$. Then we can still define the annihilation
operator $\hat a(w)$ by the formula (\ref{anni}), at least for
$\Psi\in\Gammal_\s(\cD)$.  If $w$ is unbounded, then $\hat a(w)$
is not closable and (\ref{anni1}),  the definiton of $\hat a^*(w)$ as an
operator, is incorrect.
However, we can interpret both $\hat a(w)$ and $\hat a^*(w)$ as
quadratic forms on $\Gammal_\s(\cD)$.
\label{form1}
\eer

The following inequality is sometimes called the {\em
  $N_\tau$-estimate}:
\bep Let $h>0$ and $w\in\cW$. Then
\beq \|\hat a(w)\Phi\|^2\leq(w|h^{-1}w)(\Phi|\d\Gamma(h)\Phi).\eeq
Therefore,
\beq\big\|\d\Gamma(h)^{-\frac12}\hat a^*(w)\big\|\leq\|h^{-\frac12}w\|.\eeq
\label{ntau}\eep

\proof
Clearly,
\beq |w)(w|\leq(w|h^{-1}w) h.\eeq
Applying $\d\Gamma$, we obtain
\beq \hat a^*(w)\hat a(w)=\d\Gamma\big(|w)(w|\big)\leq
(w|h^{-1}w)\d\Gamma( h).\eeq
\qed



Let $g \in \otimes_\s^2\cW.$ We define the annihilation and creation operators associated to $g$ as follows:
\begin{eqnarray}
\hat a^*(g)\Psi&:=&\sqrt{n+2}\sqrt{n+1}g\otimes_\s\Psi, \qquad
\Psi\in\otimes_\s^n\cW,\label{ja1}\\
\hat a(g)\Psi&:=&\sqrt{n+2}\sqrt{n+1} (g|\, \otimes\one^{\otimes n}\,\Psi, \qquad \Psi\in\otimes_\s^{n+2}\cW.\label{ja2}\end{eqnarray}
Again, these operators, originally defined
on $\Gamma^{{\rm fin}}_\s(\cW)$, extend to closed operators on
on $\Gamma_\s(\cW)$:

\ber Again, if $g$ does not belong to $\oplus_\s^2\cW$,
but is a functional with the domain
$\otimesal_\s^2\cD\subset\otimes_\s^2\cW $, then we can define $\hat a(g)$ 
and $\hat a^*(g)$ as quadratic forms on $\Gammal_\s(\cD)$.
\label{form2}\eer

It is important to note that each 
$g\in\otimes^2\cW$ defines a linear Hilbert-Schmidt operator from
$\bar\cW$ to $\cW$, denoted by the same symbol $g$, by the identity
\begin{eqnarray}(w_1\otimes w_2|g)&=&(w_2|g\chi w_1).\label{syme1}
\end{eqnarray}
This provides an isometric isomorphism of 
$\otimes^2\cW$ with $B^2(\bar\cW,\cW)$--the space of
Hilbert-Schmidt operators from $\bar\cW$ to $\cW$.
Symmetric tensors
(elements of  $\otimes_\s^2\cW$) are mapped onto
symmetric operators (where the  {\em symmetry} of $g$
means $g=g^\#$).

Let us state the following fact about this
identification:
\bep
Let $p_1,p_2\in B(\cW)$. Then the tensor $p_1{\otimes} p_2\, g$
corresponds to the operator $p_1 g p_2^\#$. \label{polp}\eep

\bep
Let  $w\in\cW$, $h\in B(\cW)$, $g\in  \cW\otimes_\s\cW$. Then the
following identities are true:
\begin{eqnarray}
  [\d\Gamma(h),\hat a^*(w)]=\hat a^*(hw),&&  [\d\Gamma(h),\hat
    a(w)]=-\hat a(hw),\\{}
  [\hat a(g),\hat a^*(w)]=2\hat a^*(g\bar w),&&  [\hat a^*(g),\hat
    a(w)]=-2\hat a(g\bar w).
  \end{eqnarray}
\eep

\subsection{Symplectic and metaplectic transformations
  in infinite dimensions}

As in (\ref{ess}) we introduce the operator
\beq
S=\left[\begin{array}{cc}\one&0\\0&-\one\end{array}\right].\eeq

Let $R\in B(\cW\oplus\bar\cW)$. As in Subsection \ref{Symplectic transformations},
$R$ is called symplectic
if $R^*SR=S$.
Bounded symplectic transformations form a group, which
we denote $Sp(\cY)$.

Various properties of symplectic operators described in
Subsection \ref{Symplectic transformations} are valid in the present
setting.

\bet Let 
\beq
R=\left[\begin{array}{cc}p&q\\\bar q&\bar p\end{array}\right]\in
Sp(\cY).
\label{pdf}\eeq
Then the following conditions are equivalent:
\ben\item There exists a unitary $U$ such that
\begin{eqnarray}
U\hat a^*(w)U^*&=&\hat a^*(pw)+\hat a(q\bar w),\nonumber\\
U\hat a(w)U^*&=&\hat a^*(q\bar w)+\hat a(pw),\ \ \ w\in\cW.\label{equi}
\end{eqnarray}
\item 
There exists a unitary $U$ such that
 \beq
 U\e^{\i\hat\phi(w,\bar w)}U^*=\e^{\i\hat\phi(w',\bar w')},\ \ \ \
 R\left[\begin{array}{c}w\\\bar
     w\end{array}\right]=\left[\begin{array}{c}w'\\\bar
     w'\end{array}\right],\ \ \ w\in\cW .\label{equi1}\eeq
\item There exists a $*$-automorphism $\alpha_R$ of  $B(\Gamma_\s(\cW))$ such that
 \beq
\alpha_R\big(\e^{\i\hat\phi(w,\bar w)}\big)=\e^{\i\hat\phi(w',\bar w')},\ \ \ \
R\left[\begin{array}{c}w\\\bar
     w\end{array}\right]= \left[\begin{array}{c}w'\\\bar
     w'\end{array}\right],\ \ \ w\in\cW .\eeq
 \een
Let (1), (2) and (3) be true. Then $U$ (common for (1) and (2)) is
 uniquely determined up to a phase factor. Besides, $\alpha_R$ is
 uniquely defined.\eet

 If $R$ satisfies the conditions of the above theorem, then we say
 that $R$ is {\em implementable}.
 The unitary 
$U$  is  called
 a {\em (Bogoliubov) implementer of $R$}.  $\alpha_R$ is called the {\em
   Bogoliubov automorphism associated to $R$.} 

 We leave the proof of  this theorem to the reader. Let us only mention that
 to show (3)$\Rightarrow$(2) we need to use Proposition \ref{auto1}. To
 obtain the uniqueness of $\alpha_R$ we use the weak density of
 linear combinations of Weyl operators in  $B(\Gamma_\s(\cW))$.

$Sp_\res(\cY)$ will denote the {\em restricted symplectic group},
 which consists of $R\in Sp(\cY)$ such that $q$ is Hilbert-Schmidt.
The importance of $Sp_\res(\cY)$ is due to the Shale Theorem
\cite{Sh}, which we quote below
in the form given in \cite{DG}.
\bet $R\in Sp(\cY)$. Then $R$ is implementable iff $R\in  
Sp_\res(\cY)$. For such $R$, we can define  the {\em natural implementer
  of $R$}
\begin{eqnarray}
U_R^{\nat}&:=&|\det
pp^*|^{-\frac{1}{4}}\e^{-\frac12\hat
  a^*(d_2)}\Gamma\bigl((p^*)^{-1}\bigr)\e^{\frac12\hat a(
  d_1)},\label{natural}
\end{eqnarray}
where $d_2$, $d_1$ are defined as in (\ref{dfd1}) and  (\ref{dfd2}).
All implementers of $R\in Sp_\res(\cY)$ coincide with $
U_R^\nat$, up to a phase factor.\eet

Bogoliubov implementeres form a group, which is
denoted $Mp^c(\cY)$.
We have a short exact sequence
\[\one\to U(1)\to Mp^{\rm c}(\cY)\to Sp_\res(\cY)\to\one.\]

Let us mention the following criterion, which was used in \cite{NNS}:

\bep If  $R^*R-\one$ is Hilbert-Schmidt, then $R\in Sp_\res(\cY)$.
\label{prio1}\eep
\proof
\[R^*R=\left[\begin{array}{cc}p^*p+q^\t \bar q&p^*q+q^\t \bar p
    \\ p^\t \bar q+q^*p&p^\t \bar p+q^*q\end{array}\right]
=\left[\begin{array}{cc}\one+2q^\t \bar q&2p^*q
    \\ 2p^\t \bar q&\one+2q^*q\end{array}\right].
\]
Now
\begin{eqnarray}\|\one-R^*R\|_\HS^2&=&
8\Tr q^*q+8\Tr q^*qq^*q+8\Tr q^*pp^*q\\&=&16\Tr q^*qq^*q+16\Tr q^*q\ \geq \ 
16\Tr q^*q.
\end{eqnarray}
\qed

$Sp_\af(\cY)$ will denote the {\em anomaly-free symplectic group},
which consists of $R\in Sp(\cY)$ such that $\one-p$ is trace
class \cite{DG}.
\bep
$Sp_\af(\cY)$ is a subgroup of $Sp_\res(\cY)$. \eep

\proof We have
\beq q^*q=p^*p-\one=(p^*-\one)p+p-\one.\eeq
Therefore, $\|p-\one\|_1<\infty$ implies $\|q\|_2<\infty$. \qed

For $R\in Sp_\af(\cY)$ we can define a pair of {\em metaplectic Bogoliubov implementers}
\beq
\pm U_R^\met:=
\pm (\det
p^*)^{-\frac{1}{2}}\e^{-\frac12\hat
  a^*(d_2)}\Gamma\bigl((p^*)^{-1}\bigr)\e^{\frac12\hat a( d_1)}.
\eeq
They form a group, which we denote $Mp_\af(\cY)$ \cite{DG}.
We have a short exact sequence
\[\one\to \zz_2\to Mp_{\af}(\cY)\to Sp_{\af}(\cY)\to\one.\]

\subsection{Classical quadratic Hamiltonians}

In this subsection we consider strongly continuous 1-parameter groups
of symplectic transformations. The following proposition describes their generators:

\bep Let $\i B$ be a generator of a 1-parameter group on $\cW\oplus\bar\cW$. The following statements are equivalent:
\ben\item
$\e^{\i Bt}$, $t\in\rr$, is a strongly continuous 1-parameter group of
symplectic transformations.
\item
$\i B$ is $J$-real, $SB^*\supset BS$.
\item $A_B:=BS$ is  $J$-real and $A_B^*\supset A_B$ (in other
  words, $A_B$ is Hermitian).\een  
\label{pwe}\eep

\proof
We have for $w_1,w_2\in\Dom(B)$
\beq\frac{\d }{\d t}(\e^{\i tB}w_1|S\e^{\i tB}w_2)\Big|_{t=0}
=-\i(Bw_1|Sw_2)+\i(w_1|SBw_2).\eeq
Hence preservation of $S$ by $\e^{\i tB}$ is equivalent to
$(SAS)^*=B^*S\supset SB=SAS$, which means that $SAS$ is Hermitian.
This is equivalent to $A$ being Hermitian.
\qed

For brevity, we will say that $B$ is a {\em symplectic generator} if
 $\i B$  generates  a one-parameter group of symplectic
transformations. Similarly as in the previous section, $A_B:=
BS$ will be sometimes called the {\em classical Hamiltonian of } $B$, and we will often write $A$ instead of $A_B$.

Note that in finite dimensions the converse of Proposition \ref{pwe} (3)
is true: If $A$ is
is Hermitian and $J$-real, then $B:=AS$ is a symplectic generator.
This is probably not the case in infinite dimensions.

\subsection{Bogoliubov Hamiltonians}
\label{Bogoliubov Hamiltonians}

Let  $B$  be, as usual, a symplectic generator, and $A=BS$.
We will write  \beq\e^{\i tB}=\left[\begin{array}{cc}p_t&q_t\\\bar q_t&\bar p_t\end{array}\right].\label{kala}\eeq

\bet The following conditions are equivalent:
\ben\item There exists a self-adjoint operator $\hat H$ on
$\Gamma_\s(\cW)$ such that $\e^{\i t\hat H}$ implements $\e^{\i tB}$
for any $t\in\rr$.
\item There exists  $\alpha_t$, a 1-parameter $C_0^*$-group of
  $*$-automorphisms of $B(\Gamma_\s(\cW))$, such that
\beq  \alpha_t\big(\e^{\i\hat\phi(w,\bar w)}\big)=\e^{\i\hat\phi(w_t,\bar w_t)},\ \ \ \
 \left[\begin{array}{c}w_t\\\bar
     w_t\end{array}\right]=\e^{\i tB}\left[\begin{array}{c}w\\\bar
     w\end{array}\right],\ \ \ w\in\cW .\label{equi1b}\eeq
\item
 $\lim\limits_{t\to0}\|q_t\|_\HS=0.$
 \een
 Let (1), (2), (3) be true. Then  $\alpha_t$ is determined uniquely.
 $\hat H$ is uniquely defined up to an additive
constant.
\label{haha}\eet

$\hat H$ will be called a {\em quantization of $B$.}
We will also say that $\hat H$ is a {\em quantum quadratic Hamiltonian}, or
shorter, a {\em Bogoliubov Hamiltonian}. If the equivalent conditions of the above theorem are satisfied,
then we will
say that {\em $B$ possesses  quantizations.}

\proof (1)$\Leftrightarrow$(2) is a consequence of Proposition
\ref{auto2}. We need to show that (1),(2)$\Leftrightarrow$(3)

If  $\e^{\i tB}$,  $t\in\rr$, is implementable,
then  $\|q_t\|_\HS<\infty,$  for all $t\in\rr$.

If  $\lim\limits_{t\to0}\|q_t\|_\HS=0,$ then  $\|q_t\|_\HS<\infty,$  for small enough $t$. But since $Sp_\res(\cY)$ is a group, 
$\|q_t\|_\HS<\infty,$  for all $t\in\rr$.

Thus, in all cases (1), (2) and (3)  we can define
\beq
U_t^\nat:=U_{\e^{\i tB}}^\nat,\eeq
(see (\ref{natural})). Set
\beq\alpha_t(C):=U_t^\nat CU_{-t}^\nat .\eeq
Clearly, $t\mapsto \alpha_t$ is a 1-parameter group of
$*$-automorphisms satisfying (\ref{equi1b}).
The proof will be completed if we show the equivalence
of the following statements:
\begin{romanenumerate}
\item $t\mapsto U_t^\nat$ is strongly continuous at zero;
    \item $t\mapsto\alpha_t$ is a $C_0^*$-group of
    $*$-automorphisms;
    \item $d_{2,t}:=q_t\bar p_t^{-1}$ satisfies
      $\lim\limits_{t\to0}\|d_{2,t}\|_\HS=0$;
    \item  $\lim\limits_{t\to0}\|q_{t}\|_\HS=0$.
      \end{romanenumerate}

\noindent
(i)$\Rightarrow$(ii): We easily see that
if $t\mapsto U_t^\nat$ is strongly continuous at zero and if $C$ is a
bounded operator, then
$t\mapsto U_t^\nat CU_{-t}^\nat $ is weakly continuous at zero. This
implies that $t\mapsto\alpha_t$ is a $C_0^*$-group.

\noindent(ii)$\Rightarrow$(iii): Let $|\Omega)(\Omega|$ denote the
orthogonal projection onto $\Omega$. We have
\begin{eqnarray}
  \Big(\Omega|\alpha\big(|\Omega)(\Omega|\big)\Omega\Big)&=&
  \big|(\Omega|U_t^\nat\Omega)\big|^2\nonumber\\
  =\ \big|\det p_tp_t^*\big|^{-\frac12}&=&
  \det\big(\one- d_{2,t}^* d_{2,t}\big)^{\frac12}\label{pro0}\\
  &=&\exp\Big(\frac12\Tr\log\big(\one-d_{2,t}^*d_{2,t}\big)\Big).
  \label{pro}\end{eqnarray}
  In (\ref{pro0}) we used the identity
  \beq p_t^{\t-1}\bar p_t^{-1}=\one-d_{2,t}^*d_{2,t}.\label{quao}
  \eeq
(ii) implies that (\ref{pro}) goes to $1$ for $t\to0$. This is
  equivalent to $\lim\limits_{t\to0}\Tr \log(\one-d_{2,t}^*
  d_{2,t})=0$, which is equivalent to
  $\lim\limits_{t\to0}\Tr d_{2,t}^* d_{2,t}=0$.

\noindent(iii)$\Rightarrow$(i): We have
\beq U_t^\nat\e^{\i\hat\phi(w,\bar w)}\Omega
=\e^{\i\hat\phi(w_t,\bar w_t)}|\det
p_tp_t^*|^{-\frac{1}{4}} \e^{-\frac12a^*( d_{2,t})}\Omega.\label{pro1}\eeq
But $t\mapsto \e^{\i\hat\phi(w_t,\bar w_t)}$ is strongly continuous.
By (\ref{quao}), $\lim\limits_{t\to0}|\det
p_tp_t^*|^{-\frac{1}{4}}=1$.  
Besides, (iii) implies that $\lim\limits_{t\to0} \e^{-\frac12a^*(
  d_{2,t})}\Omega=\Omega$.
Therefore, (\ref{pro1})  is continuous at $t=0$.
But the span of $\e^{\i\hat\phi(w,\bar w)}\Omega$ is dense and
$ U_t^\nat$ is unitary. Hence $ U_t^\nat$ is strongly continuous at
$t=0$.

\noindent(iii)$\Leftrightarrow$(iv) follows from the identities
\begin{eqnarray}
  q_tq_t^*&=&d_{2,t}^*d_{2,t}\big(\one-d_{2,t}^*d_{2,t}\big)^{-1},\\
  d_{2,t}^*d_{2,t}&=& q_tq_t^*\big(\one+ q_tq_t^*\big)^{-1}.
  \end{eqnarray}
\qed


 Below we describe 3 distinguished
quantizations.

\ben\item
If the group $\e^{\i t\hat H}$ implementing $\e^{\i tB}$
is contained in $Mp_\af(\cY)$, then 
$\hat H$ will be called  {\em Weyl}.
It is easy to see that for a given symplectic generator $B$, its {\em Weyl
quantization}, if it exists, is unique. We will denote it
 by $\hat H_B^\w$. An alternative name for  $\hat H_B^\w$:
the {\em symmetric quantization of $B$}.
\item
We say that a quantization  $\hat H$ of $B$  is  {\em normally ordered}
\beq\frac{\d}{\d t}(\Omega|\e^{\i t\hat H}\Omega)\Big|_{t=0}=0.
\label{qua4}\eeq
Again, a given symplectic generator $B$ possesses at most one {\em normally
ordered quantization}. We  will denote it by $\hat H_B^\n$.
An alternative name for $\hat H_B^\n$: the {\em Wick quantization of $B$}.
\item
If $B$ possess a quantization, which is bounded from below, then all
of its quantizations are bounded from below. Then one can introduce
the {\em zero-infimum quantization} $\hat H_B^\z$ fixed by
the condition
\[\inf\hat H_B^\z=0.\]
\een

Let us stress that there exist $B$ that possess  quantizations, but
they do not possess $\hat H_B^\w$, $\hat H_B^\n$ or $\hat H_B^\z$.

We will usually drop the subscript $B$ in the above symbols.

Note that whereas the definitions of $\hat H^\w$
and $\hat H^\z$ are quite obvious,  it is less clear how to
generalize the concept of normally ordered Bogoliubov Hamiltonian to
infinite dimensions. In the following proposition we formulate another
condition, which could be  considered as another candidate
for a definition of $\hat H^\n$.

\bep Suppose that $B$ possesses a quantization $\hat H$ such that
$\Omega\in\Dom\big(|\hat H|^{\frac12}\big)$ (the vacuum belongs to
the form domain of $\hat H$). Then $B$ possesses the normally ordered
quantization.\label{normal}
\eep

\proof We easily check that
\beq \hat H^\n:=\hat H-(\Omega|\hat H\Omega).\label{normal1}\eeq
 satisfies (\ref{qua4}). \qed

\bet Consider (\ref{kala}).
\ben \item The condition  \beq\lim_{t\to0}\|p_t-\one\|_1=0.\label{qua6}\eeq
is equivalent to  $B$ possessing the
Weyl quantization $\hat H^\w$. If this is the case, then
\beq 
 \e^{\i t\hat H^\w}=
(\det
p_t^*)^{-\frac{1}{2}}\e^{-\frac12\hat
  a^*(d_{2,t})}\Gamma\bigl((p_t^*)^{-1}\bigr)\e^{\frac12\hat a( d_{1,t})},
\label{qua1}
\eeq
where the sign of the square root is determined by continuity.

\item
  Suppose that there exists a self-adjoint operator $h$ on $\cW$ such
  that
  \beq\lim_{t\to0}\frac{\|\e^{-\i th} p_t-\one\|_1}{t}=0.\label{qua3}\eeq
  Then $B$ possesses the normally ordered quantization $\hat H^\n$ and
  \beq
 \e^{\i t\hat H^\n}=
(\det
p_t^*\e^{\i th})^{-\frac{1}{2}}\e^{-\frac12\hat
  a^*(d_{2,t})}\Gamma\bigl((p_t^*)^{-1}\bigr)\e^{\frac12\hat a( d_{1,t})},
\label{qua2}\eeq
where the sign of the square root is determined by continuity.
The operator $h$ that appears in (\ref{qua3}) is uniquely defined.
\item Suppose that the assumptions of (2) hold. In addition, assume
  that
  $h$  in (\ref{qua3}) is trace class.
  Then $B$ possesses both normally ordered and Weyl quantization, and
  \beq \hat H^\n+\frac12\Tr h=\hat H^\w.\eeq
  \een\label{kala1}
  \eet

\proof (1): $\lim\limits_
{t\to0}\|\one-p_t\|_1=0$ implies that $\e^{\i tB}\in
Sp_\af(\cY)$ at least for small $t$. But $Sp_\af(\cY)$ is a group. Therefore,
$\e^{\i tB}\in Sp_\af(\cY)$ for all $t\in\rr$.

Besides, $t\mapsto \e^{\i tB}$ is continuous in the topology of $Sp_\af(\cY)$ at zero. By the group property of $Sp_\af(\cY)$, it is continuous for all $t\in \rr$.

Hence,
$U_t^\met$ given by (\ref{qua1}) is a well defined. $U_t^\met$ obviously is
one of the metaplectic  implementers of $\e^{\i
  tB}$. We have \beq(\Omega|U_t^\met\Omega)=(\det
p_t^*)^{-\frac{1}{2}}\label{quau}\eeq
depends continuously on $t$.

Using that $Mp(\cY)$ is a group and the continuity of
(\ref{quau}), we see that  $U_t^\met$
satisfies the
group property.
Next, repeating the argument of the proof of Thm \ref{haha},
we see that  $U_t^\met$ is contiunuous on coherent vectors.

Thus  $U_t^\met$ is a strongly continuous group of metaplectic implementers of 
$\e^{\i tB}$. Hence $B$ posseses the Weyl quantization.

Conversely, if $B$ possesses the Weyl quantization $\hat H^\w$, then
$U_t^\met=\e^{\i t\hat H^\w}$.
Then (\ref{quau}) is true. But
$\lim\limits_
{t\to0}\|\one-p_t\|_1=0$ is equivalent to the continuity of the rhs of (\ref{quau}).

(2):
(\ref{qua3}) implies
\beq
\lim\limits_{t\to0}\|\e^{-\i th} p_t-\one\|_1=0.\label{das}\eeq
Therefore, the identity
\beq q_t^*q_t=p_t^*p_t-\one=(p_t^*\e^{\i th}-\one)\e^{-\i
  th}p_t+\e^{-\i th}p_t-\one\eeq
shows that $\lim\limits_{t\to0}\|q_t\|_2=0$. Therefore,
(\ref{qua2}) is  well defined and depends continuously on $t$.
Clearly,
\beq\big|\det p_t^*\e^{\i th}\big|^2=\det p_t^*p_t.\eeq
Hence, (\ref{qua2}) differs from $U_{\e^{\i
    tB}}^\nat$ by a phase factor. We check by
direct calculation that it satisfies the group property
\cite{BD}. 

Using (\ref{das}) and the differentiability of the determinant in the
trace norm, for small enough $t$ we have
\beq\big|\det p_t^*\e^{\i th}-1\big|\leq c\|p_t^*\e^{\i
  th}-\one\|_1.\eeq
Hence (\ref{qua3}) implies
\beq\lim_{t\to0}\frac{\det
  p_t^*\e^{\i th}-1}{t}=0.\eeq
Therefore, using also (\ref{das}),
\beq\lim_{t\to0}\frac{(\det
  p_t^*\e^{\i th})^{-\frac12}-1}{t}=0.\eeq
By (\ref{qua2}),
\beq \big(\Omega| \e^{\i t\hat H^\n}\Omega\big)=
(\det
p_t^*\e^{\i th})^{-\frac{1}{2}}.\eeq
Hence, (\ref{qua4}) is true.

Suppose that for $h_1,h_2$ we have (\ref{qua3}). Let $w,w'\in\cW$ be normalized. Then
\begin{eqnarray}
  \frac1t\big|(w|\e^{\i th_1}w')-(w|\e^{\i th_2}w')\big|&\leq&
  \frac1t \|\e^{\i th_1}-\e^{\i th_2}\|\\
  &\leq&\frac1t \|\e^{\i th_1}-\e^{\i th_2}\|_1\\
  &\leq&\frac1t \|\e^{\i th_1}-p_t\|_1+\frac1t\|p_t-\e^{\i th_2}\|_1\ \to\ 0.
\end{eqnarray}
Hence  $h_1=h_2$ by Lemma \ref{qua8}.

(3):
Using $\|h\|_1<\infty$, we can write
\beq\det p_t=\det\e^{\i th}\det\e^{-\i th}p_t=
\e^{\i t\Tr h}\det\e^{-\i th}p_t.\eeq
Thus we see that both (\ref{qua1}) and (\ref{qua2}) are well defined
and
\beq \e^{\i t\hat H^\w}=\e^{\i t\frac12\Tr h}\e^{\i t\hat H^\n}.\eeq
\qed

\subsection{Criteria for existence of quantizations of classical Hamiltonians}
\label{Bogoliubov Hamiltonians-1}

In this subsection we
restrict our study to symplectic generators that are
bounded perturbations of diagonal symplectic generators.

We will always assume that $h$ is a
self-adjoint operator on $\cW$ and  $g=g^\t$. Besides,
\begin{eqnarray}
B:=\left[\begin{array}{cc}h&-g\\\bar g&-\bar h\end{array}\right],&&
B_0:=\left[\begin{array}{cc}h&0\\0&-\bar
    h\end{array}\right],\label{hamu1b}\\
A=BS=\left[\begin{array}{cc}h&g\\\bar g&\bar h\end{array}\right],&&
A_0=B_0S=\left[\begin{array}{cc}h&0\\0&\bar
    h\end{array}\right],\ \ G=\left[\begin{array}{cc}0&-g\\\bar g&0\end{array}\right].\label{hamu1a}
\end{eqnarray}

The following proposition is immediate:

\bep If $g$ is bounded, then $B$ is a symplectic generator.
Besides, $A_B=BS$ is self-adjoint. \label{mammu}
\eep

\proof Clearly, $B_0$ is a symplectic generator and $A_0$ is self-adjoint.
We can add a bounded perturbation without destroing these properties. \qed

The following theorem is a slightly strengthened version of a
criterion  due to Berezin \cite{Be}, see also \cite{BD}.
 Throughout the subsection we set
\beq f(t):=\int_0^t\e^{\i sh}g\e^{\i s\bar h}\d s.\eeq

\bet
\ben\item
Suppose that $g$ is bounded and  $\lim\limits_{t\to0}\|f(t)\|_2=0$. Then $B$ possesses quantizations.
\item In addition to assumptions of (1) suppose that
  $\lim\limits_{t\to0}\|\bar gf(t)\|_1=0$. Then $B$ possesses the normally ordered
  quantization.
  \item In addition to assumptions of (2) suppose that
    $\|h\|_1<\infty$. Then $B$ possesses
     both the Weyl and the normally ordered  quantizations, and
  \beq \hat H^\n+\frac12\Tr h=\hat H^\w.\eeq \een
\label{posi3}\eet

\proof (1):
Using repeatedly the identity
\beq f(2t)=f(t)+\e^{\i th}f(t)\e^{\i t\bar h}\eeq
we see that $\|f(t)\|_2$ is finite for all $t$.

Set
\begin{eqnarray}
  V(t)&:=&\e^{\i tB}\e^{-\i tB_0},\\
   G(t)&:=&\e^{\i tB_0}G\e^{-\i
     tB_0}\ =\ \left[\begin{array}{cc}0&-\e^{\i sh}g\e^{\i s\bar h}
       \\\e^{-\i s\bar h}\bar g\e^{-\i s h}
       &0\end{array}\right],\\
   F(t)&:=&\int_0^tG(s)\d s\ =\ \left[\begin{array}{cc}0&-f(t)
       \\\bar{f(t)}
       &0\end{array}\right]
   .\end{eqnarray}
From
\beq V(t)=\one+\i\int_0^tV(s)G(s)\d s\label{posi2}\eeq
and $\|G(t)\|=\|G\|$, we obtain
\beq \|V(t)\|\leq\e^{|t|\|G\|}.\eeq
  Iterating (\ref{posi2}) gives
  \beq
  V(t)=\one+\i F(t)-\int_0^tV(s)G(s)\e^{\i sB_0}F(t-s)\e^{-\i sB_0}\d
  s.\label{qui1}
  \eeq
  Therefore,
  \beq\|V(t)-\one\|_\HS\leq
  \|F(t)\|_\HS+\int_0^t\|V(s)\|\|G\|\|F(t-s)\|_\HS\d s.\eeq
  But $\|F(t)\|_\HS=\sqrt2\|f(t)\|_\HS$, $\|G\|=\|g\|$.
  Hence $\|V(t)-\one\|_\HS$ is finite and
   goes to zero
   as $t\to0$. Arguing as in Proposition \ref{prio1}, we obtain
\beq 16\|q(t)\|_\HS^2\leq\|V(t)-\one\|_\HS.\eeq
   Therefore, $\|q(t)\|_\HS$ is finite   and
   goes to zero
   as $t\to0$. This means that the assumption of
   Theorem \ref{haha} (3) is satisfied. Hence
$B$ possesses quantizations.

   (2): We rewrite (\ref{qui1}) as
   \begin{eqnarray}
     \left[\begin{array}{cc}p_t-\e^{\i th}&q_t+\i f(t)\e^{-\i t\bar h}
       \\\bar q_t-\i \bar{f(t)}\e^{\i th}
       &\bar p_t-\e^{-\i th}\end{array}\right]
     &=&\e^{\i tB}-\e^{\i
       tB_0}-\i F(t)\e^{\i tB_0}\\
     &=&-\int_0^tV(s)G(s)\e^{\i sB_0}F(t-s)\e^{\i(t- s)B_0}\d
  s.\label{qui2}
\end{eqnarray}
Therefore, by (\ref{traceclass}),
   \begin{eqnarray}2\|p_t-\e^{\i th}\|_1&\leq &\|\e^{\i tB}-\e^{\i
       tB_0}-\i F(t)\e^{\i tB_0}\|_1\\
     &\leq&
        \int_0^t\|V(s)\|\|GF(t-s)\e^{\i(t- s)B_0}\|_1\d
  s.\label{qui3}
\end{eqnarray}
   Using  $\|GF(t)\|_1=2\|\bar gf(t)\|_1$ and
    $\lim\limits_{t\to0}\|\bar gf(t)\|_1=0$, we see that (\ref{qui3}) is
$o(t)$. Thus we obtain $\|p_t-\e^{\i th}\|_1=o(t)$. This means that the
   assumption of Theorem \ref{kala1} (2) is satisfied. Hence, $B$
   possesses the normally ordered quatization.

   (3):  We apply  Theorem \ref{kala1} (3).
   \qed

The assumptions of   Theorem \ref{posi3} are not very convenient to verify. Our next aim is to formulate  criteria for the existence of quantizations, which are more conveneient to check.

Define
  \beq \gamma(g):=(h\otimes\one+\one\otimes h)^{-1}g,\eeq
  where we use the tensor interpretation of $g$ and assume that
  $g\in\Dom(h\otimes\one+\one\otimes h)^{-1}$.
\bep 
  In the operator
  interpretation, $\gamma(g)$ corresponds to
  \beq
  \gamma( g)=\i\lim_{\epsilon\searrow0}\int_0^\infty\e^{-\epsilon t}
  \e^{-\i th}g\e^{- \i t\bar h}\d t
  \label{wick}\eeq
  and satisfies
  \begin{eqnarray}
    h\gamma(g)+\gamma(g)\bar h&=&g.\label{wick2}\end{eqnarray}
For $h>0$ we can ``Wick rotate'' the 
formula (\ref{wick}) and write
  \beq
  \gamma( g)=\int_0^\infty
  \e^{- th}g\e^{-t\bar h}\d t.\label{wick1}\eeq\eep

  \proof By Prop. \ref{polp} and
  $\bar h=h^\#$, we can identify the operator $\e^{-\i th}g\e^{- \i
    t\bar h}$ with
  the tensor
  \beq \e^{-\i th}\otimes\e^{-\i th} g
  =\e^{-\i t(h\otimes\one+\one\otimes h)} g.\eeq
  Clearly,
  \beq
  \i\int_0^\infty\e^{-\epsilon t}\e^{-\i t(h\otimes\one+\one\otimes
    h)} g
  =(h\otimes\one+\one\otimes
  h-\i\epsilon)^{-1}g\to
  (h\otimes\one+\one\otimes h)^{-1}g,\eeq
where we use the usual Hilbert space convergence,   which proves (\ref{wick}).

Set
\beq\gamma_\epsilon(g):=\i\int_0^\infty\e^{-\epsilon s}
\e^{-\i sh}g\e^{- \i s\bar h}\d s.\eeq
We compute:
\begin{eqnarray}
  \e^{-\i th}\gamma_\epsilon(g)\e^{-\i t\bar h}&=&
\i\int_t^\infty\e^{-\epsilon (s-t)}
\e^{-\i sh}g\e^{- \i s\bar h}\d s\\&=&
  -\i\int_0^t\e^{-\i
  sh}g\e^{-\i s\bar h}\e^{-\epsilon s}\d s+\e^{\epsilon
    t}\gamma_\epsilon(g).
  \end{eqnarray}
We differentiate wrt $t$ at $t=0$, obtaining
\beq -\i\big(h\gamma_\epsilon(g)+\gamma_\epsilon(g)\bar h\big)=-\i
g+\epsilon\gamma_\epsilon(g).\eeq
Taking the limit as $\epsilon\searrow0$, we obtain (\ref{wick2}).

  The proof of (\ref{wick1}) is almost the
same as that of (\ref{wick}). \qed

  We will also write
\beq  \gamma(G):=\left[\begin{array}{cc}0&-\gamma(g)\\\bar{\gamma(g)}&0\end{array}\right].
\eeq
Note that in the operator interpretation we have
\begin{eqnarray}
 [B_0, \gamma(G)]&=&G.\end{eqnarray}

The following criterion is a consequence
of
Theorem \ref{posi3}.
\bet\ben\item Suppose that $g$ is bounded and $g=g_1+g_2$, where
$\|g_1\|_\HS<\infty$ and $\|
\gamma(g_2)\|_\HS<\infty$. Then the assumptions of Theorem \ref{posi3} are
satisfied, and hence $B$ possesses quantizations.
\item Suppose that $\|g\|_\HS<\infty$. Then $B$ possesses the normally
  ordered quantization.
  \item Suppose that $\|h\|_1<\infty$ and  $\|g\|_\HS<\infty$. Then
    $B$ possesses both the Weyl and the normally
    ordered quantization. Besides,
    \beq \hat H^\w=\hat H^\n+\Tr h.\eeq\een
\label{th2}\eet

\proof
(1): Set
\beq f_i(t):=\int_0^t\e^{\i sh}g_i\e^{\i s\bar h}\d s.\eeq
It is clear that  $\lim\limits_{t\to0}\|f_1(t)\|_2=0$.
The fact that  $\lim\limits_{t\to0}\|f_2(t)\|_2=0$
 follows from
\beq 
\int_0^t\e^{\i s h}g_2\e^{\i s \bar h}\d s=
-\i\int_0^t\frac{\d}{\d s}\e^{\i s h}\gamma(g_2)\e^{\i s \bar h}\d s=
-\i\e^{\i t h}\gamma(g_2)\e^{\i t \bar h}+\i \gamma(g_2).\eeq
Hence assumptions of Thm  \ref{posi3} (1) are satisfied.

(2): Clearly,  $\|\bar g f(t)\|_1\leq t\|g\|_\HS^2$. Hence assumptions of
Thm \ref{posi3} (2) are satisfied.

Now (3) follows immediately from Thm \ref{posi3} (3).
\qed

Note that the assumption of Theorem \ref{hanuj} implies that of
Theorem \ref{th2}:
\bep $h>0$ and
$\|h^{-\frac12}g\bar h^{-\frac12}\|_\HS<\infty$ implies $\|\gamma(g)\|_\HS<\infty$. \eep

\proof
$h^{-\frac12}g\bar h^{-\frac12}$ corresponds to
$h^{-\frac12}\otimes h^{-\frac12} g$ in the tensor interpretation.
Clearly,
\beq 2h\otimes h\leq(h\otimes\one+\one\otimes h)^2.
\eeq
Hence,
\beq h^{-1}\otimes h^{-1}\geq 2 (h\otimes\one+\one\otimes h)^{-2}.\eeq
Therefore,
\beq
\|h^{-\frac12}g\bar h^{-\frac12}\|_\HS=
\|h^{-\frac12}\otimes h^{-\frac12} g\|\geq\sqrt{2}
\| (h\otimes\one+\one\otimes h)^{-1}g\|=\sqrt{2}  \|\gamma(g)\|_\HS
.\eeq
\qed

\subsection{Positive classical Hamiltonians and their diagonalization}

The following theorem is an extension of Theorem \ref{th1} to
arbitrary dimensions. It says that a large class of classical
Hamiltonians
can be diagonalized by a positive symplectic
transformation. This theorem is implicitly contained in \cite{DG} (see Thm
11.20 (3) together with Thm 18.5 (3)). \cite{NNS} contains also a related
result about the diagonalizabilty of classical Hamiltonians. It does
not provide, however, a construction of a distinguished diagonalizing operator.

We will use the notation introduced in (\ref{hamu1b}) and (\ref{hamu1a})
We will  assume that $h>0$. It will not be necessary to
assume that $g$ is bounded---we will assume that $g=g^\t$ is a bilinear form
with the right domain
$\Dom |\bar h|^{\frac12}$ and the left domain $\Dom | h|^{\frac12}$.

\bet
Let $h$ be positive and
  \beq\| h^{-\frac12}g\bar h^{-\frac12}\|=:a<1.\label{assu1}\eeq
  Then $A$ is a positive self-adjoint operator
  with the form domain  $\Dom A_0^{\frac12}$. The coresponding $B$ is
  a symplectic generator.

  Besides,
 \beq
R_0=SA^{-\frac12}(A^{\frac12}SAS A^{\frac12})^{\frac12}A^{-\frac12}S,
\label{infi3-}\eeq
is a bounded invertible positive symplectic operator. Hence so is
\beq
R=R_0^{\frac12}
\label{infi3d}\eeq
$R$
diagonalizes $B$ and $A$, that
is, for some positive self-adjoint $h_\dg$
\begin{align}
B&=R\left[\begin{array}{cc}h_\dg&0\\0&-\bar{
      h_\dg}\end{array}\right]R^{-1},\label{hamuj9}\\
A&=R\left[\begin{array}{cc}h_\dg&0\\0&\bar{
      h_\dg}\end{array}\right]R^*.\label{hamuj9-}
\end{align}
Moreover,
\beq\frac{(1-a)^{\frac14}}{(1+a)^{\frac14}}\leq\|R\|
\leq \frac{(1+a)^{\frac14}}{(1-a)^{\frac14}}.\eeq
\label{hamuj}\eet

\proof
$GS$ is a form bounded perturbation of $A_0$:
\[|(v|GSv)|\leq a(v|A_0v),\ \ \ v\in\Dom(A_0^{\frac12}).\]
Therefore, $A$ extends to a positive self-adjoint operator by the KLMN
Theorem.

$A$ satisfies
\beq
(1-a)A_0\leq A\leq (1+a)A_0.\eeq
Similarly,
\[SAS=A_0-GS\]
extends to a positive operator satisfying
\beq
(1-a)A_0\leq SAS\leq (1+a)A_0.\eeq
Therefore
\beq A^{\frac12}SAS A^{\frac12}\geq
(1-a)A^{\frac12}A_0 A^{\frac12}\geq
\frac{(1-a)}{(1+a)}A^2.
\label{hamub}\eeq
\beq A^{\frac12}SAS A^{\frac12}\leq
(1+a)A^{\frac12}A_0 A^{\frac12}\leq
\frac{(1+a)}{(1-a)}A^2.
\label{hamub0}\eeq
Hence,
\beq \frac{\sqrt{1-a}}{\sqrt{1+a}}A\leq (A^{\frac12}SAS A^{\frac12})^{\frac12}\leq
\frac{\sqrt{1+a}}{\sqrt{1-a}}A.
\label{hamub1}\eeq

Thus $R_0$, defined by (\ref{infi3-}),
is a well defined bounded invertible
positive operator.  Hence so is $R$.

Repeating the arguments of the proof of Theorem \ref{th1}, we obtain (\ref{hamuj9}) and
(\ref{hamuj9-}). By (\ref{hamuj9}) we have
\beq
\e^{\i tB}=R\left[\begin{array}{cc}\e^{\i th_\dg}&0\\0&\e^{-\i t\bar{
        h_\dg}}\end{array}\right]R^{-1}.\label{hamuj9+}
\eeq
(\ref{hamuj9+}) is clearly symplectic. Hence $B$ is a symplectic generator.
\qed

For further use we note that we can rewrite (\ref{infi3-}) as follows:
\begin{eqnarray*}
R_0&=&SA^{-\frac12}\Big(\int\frac{\tau^2}{(\tau^2+A^{\frac12}SASA^{\frac12})}\frac{\d
  \tau}{2\pi}\Big)
A^{-\frac12}S
.\end{eqnarray*}

As a side remark note that $h>0$ and  (\ref{assu1}) not only imply
$A>0$, but the converse implication is ``almost true''.
More precisely, set $\cW_0:=(\Ker h)^\perp$.
Then $A\geq0$ is equivalent to the
following condition:
\ben \item $h\geq0$, \item
$\Ker g\supset\cW_0^\perp$ (and hence, since $g= g^\#$, we have $\Ran
g\subset\cW_0$)
\item $\|h^{-\frac12}g\bar h^{-\frac12}\|\leq1$, in the sense of operators
  from $\cW_0$.
  \een

\subsection{Implementable diagonalizability of positive Hamiltonians}

The following theorem is due to \cite{NNS}. The proof that we present
below follows closely that of \cite{NNS}, with only minor modifications.

\bet In addition to the assumptions of Thm \ref{hamuj}, suppose that
  \begin{eqnarray}
\| h^{-\frac12} g\bar h^{-\frac12}\|_\HS&<&\infty.
  \end{eqnarray} Let $R$ be the symplectic operator given by Thm
  \ref{hamuj} and $q$ be
  given by (\ref{pdf}). Then
  \beq\|q\|_\HS\leq 2\frac{1}{(1-a)}\| h^{-\frac12} g\bar
  h^{-\frac12}\|_\HS.\label{hanuj1} \eeq
  In particular, $R\in Sp_\res(
  \cY)$ and hence $R$ is implementable.
  \label{hanuj}\eet
  
  Under the assumptions of the above theorem, $R$ possesses
  a Bogoliubov implementer $U$. If $h_\dg$ is given by (\ref{hamuj9}),
  then 
\beq U\d\Gamma(h_\dg)U^*\label{posa}\eeq
is the zero-infimum quantization of $B$.  (\ref{posa}) possesses a
ground state (its infimum is an eigenvalue). 

\medskip

  \noindent
{\bf Proof of Thm \ref{hanuj}.}
  We start from estimating
$R^{*}R-\one=R^2-\one=R_0-\one$. 
  We have
  \begin{eqnarray*}
    &&    S(R_0-\one)S\\
    &=&A^{-\frac12}(A^{\frac12}SASA^{\frac12})^{\frac12} A^{-\frac12}-
    A^{-\frac12}( A^2)^{\frac12} A^{-\frac12}\\
    &=&
    \int\frac{\d\tau}{2\pi} A^{-\frac12}
    \Big(\frac{A^{\frac12}SASA^{\frac12}}
             {\tau^2+    A^{\frac12}SASA^{\frac12}}-
             \frac{A^2}{\tau^2+A^2}\Big) A^{-\frac12}\\
                 &=&
-    \int\frac{\tau^2\d\tau}{2\pi} A^{-\frac12}
    \Big(\frac{1}
             {\tau^2+    A^{\frac12}SASA^{\frac12}}-
             \frac{1}{\tau^2+A^2}\Big) A^{-\frac12}\\
                 &=&
    \int\frac{\tau^2\d\tau}{2\pi}
    A^{\frac12} \frac{1}
    {\tau^2+    A^{\frac12}SASA^{\frac12}} A^{-\frac12}
    (SAS-A)
    \frac{1}{\tau^2+A^2}\\
    &=:&\int\frac{\tau^2\d\tau}{2\pi} T(\tau).
                    \end{eqnarray*}
  Now, for any $\epsilon>0$,
  \begin{eqnarray*}
&&   \pm 2T(\tau)\\
&=& \pm 2A^{-\frac12} \frac{1}
    {\tau^2+    A^{\frac12}SASA^{\frac12}} A^{\frac12}
    (SAS-A)
    \frac{1}{\tau^2+A^2}\\
    &\leq&\epsilon^{-1}    \frac{1}{\tau^2+A^2}
    (SAS-A)
     A^{\frac12} \frac{1}
    { A^{\frac12}SASA^{\frac12}} A^{\frac12}
    (SAS-A)
     \frac{1}{\tau^2+A^2}\\
     &&+\ \ \epsilon
      A^{-\frac12} \frac{ A^{\frac12}SASA^{\frac12}}
    {\tau^2+    A^{\frac12}SASA^{\frac12}} A^{-\frac12}
\\&=:&
\epsilon^{-1}T_1(\tau)+\epsilon T_2(\tau).
                    \end{eqnarray*}
We deal with the second term:
  \begin{eqnarray*}
    &&    \int\frac{\tau^2\d\tau}{2\pi} T_2(\tau)\\
    &=&\int\frac{\tau^2\d\tau}{2\pi}
      A^{-\frac12} \frac{ A^{\frac12}SASA^{\frac12}}
      {\tau^2+    A^{\frac12}SASA^{\frac12}} A^{-\frac12}\\
      &=&
 A^{-\frac12} ( A^{\frac12}SASA^{\frac12})^{\frac12}
 A^{-\frac12}\\
 &=&R_0.
  \end{eqnarray*}
  Next we treat the first term:
\begin{eqnarray*}
    &K:=&    \int\frac{\tau^2\d\tau}{2\pi} T_1(\tau)\\
    &=&  \int\frac{\tau^2\d\tau}{2\pi}
 \frac{1}{\tau^2+A^2}
    (SAS-A)
    S \frac{1}
    { A}S
    (SAS-A)
     \frac{1}{\tau^2+A^2}.
  \end{eqnarray*}
We have,
\begin{eqnarray}
&\Tr K\
    =&\int\frac{\tau^2\d\tau}{2\pi}
       \frac{1}{(\tau^2+A^2)^2}
    (SAS-A)
     S \frac{1}
    { A}S
    (SAS-A)\\
    &=&\Tr \frac{1}{A}
    (SAS-A)
     S\frac{1}
    { A} S
    (SAS-A)\\
    &\leq&\frac{1}{(1-a)^2}\Tr \frac{1}{A_0}
    (SAS-A)
     \frac{1}
    { A_0} 
    (SAS-A)\\
           &=&8\frac{1}{(1-a)^2}\Tr\bar h^{-1}\bar g h^{-1}g.\label{bound}
\end{eqnarray}
Thus we proved that
\beq
\pm2S(R_0-\one)S\leq\epsilon^{-1}K+\epsilon SR_0S,\label{pqu}\eeq
where $K$ is positive  operator with  trace bounded by (\ref{bound}).
We rewrite (\ref{pqu}) with sign $+$ as
\beq (2-\epsilon)S(R_0-\one)S\leq\epsilon^{-1}K+\epsilon. \label{pqu1}\eeq

Let $s_n(C)$ denote the $n$th singular value of an operator $C$, that
means, the $n$th eigenvalue of $|C|:=\sqrt{C^*C}$ in the descending
order. We will write for brevity $\lambda_n:=s_n(|q|^2)$.

Using $U$ defined in (\ref{propo}), we have
\beq
R_0-\one=R^2-\one=
2U\left[\begin{array}{cc}u(|q|^2+|q|\sqrt{\one+|q|^2})u^*
    &0\\0&|q|^2-|q|\sqrt{\one+|q|^2}
  \end{array}\right]U^*.\eeq
Therefore,
\beq s_n(R_0-\one)=2(\lambda_n+\sqrt{\lambda_n+\lambda_n^2}).\eeq
Thus,
\beq2(2-\epsilon)(\lambda_n+\sqrt{\lambda_n+\lambda_n^2})
\leq\epsilon^{-1}s_n(K)+\epsilon
.\label{cle1}\eeq

Let $c$ be an arbitrary positive  number. Let \beq
\lambda_n\leq c.\label{cla1}\eeq
Clearly, 
(\ref{cle1}) implies
\beq2(2-\epsilon)\sqrt{\lambda_n}\leq\epsilon^{-1}s_n(K)+\epsilon.\eeq
Taking into account (\ref{cla1}), we obtain
\beq4\sqrt{\lambda_n}\leq\epsilon^{-1}s_n(K)+\epsilon(1+2\sqrt{c}).\eeq
Optimizing wrt $\epsilon$, we obtain
\beq4\sqrt{\lambda_n}\leq2\sqrt{s_n(K)}\sqrt{1+2\sqrt{c}}.\eeq
Hence,
\beq\lambda_n\leq s_n(K)\frac{1+2\sqrt{c}}{4}.\label{clu1}\eeq

Let
 \beq
c\leq \lambda_n.\label{cla2}\eeq
Clearly, 
(\ref{cle1}) implies
\beq
4(2-\epsilon)\lambda_n\leq\epsilon^{-1}s_n(K)+\epsilon.\eeq
Taking into account (\ref{cla2}), we obtain
\beq
\lambda_n\leq\frac{1}{\epsilon\big(8-\epsilon(4+c^{-1})\big)}s_n(K).\eeq
Optimizing wrt $\epsilon$, we obtain
\beq\lambda_n\leq\frac{4+c^{-1}}{16}s_n(K).\label{clu2}\eeq

Setting $c=\frac14$ in (\ref{clu1}) and (\ref{clu2}), we obtain
\beq\lambda_n\leq\frac12s_n(K).\label{clu0}\eeq
Hence,
\beq\|q\|_\HS^2=\sum_{n=1}^\infty\lambda_n\leq \frac12\sum_{n=1}^\infty s_n(K)=\frac12\Tr K.\eeq
This together with (\ref{bound}) yields (\ref{hanuj1}).
\qed


\subsection{Normally ordered Hamiltonian}
\label{Normally ordered Hamiltonian}

In this subsection we give conditions on $B$ that guarantee the existence of a
bounded from below normally ordered quantization. We follow
\cite{NNS}, whose approach is  based on
quadratic forms. Similar results were contained in
\cite{BD}. They were however weaker, since only
 operator bounded perturbations were used in \cite{BD}.

Suppose that $\Phi,\Psi\in\Gamma_\s(\cW)$. Define the {\em reduced 1-body
density operator} $\gamma_{\Psi,\Phi}$ and the {\em pairing operator}
$\alpha_{\Psi,\Phi}$ as follows:
\begin{eqnarray*}
  (w_1|\gamma_{\Psi,\Phi}w_2)&:=&(\Phi|\hat a^*(w_2)\hat a(w_1)\Psi),\\
  (\alpha_{\Phi,\Psi} \bar w_2|w_1)=(\alpha_{\Phi,\Psi}|w_1\otimes
  w_2)&:=&(\Phi|\hat a^*(w_2)\hat a^*(w_1)\Psi),\ \ \ \ w_1,w_2\in\cW.
\end{eqnarray*}
(Note that, as usual for similar objects, $\alpha_{\Psi,\Phi}$ has two
interpretations: as a symmetric operator from $ \bar\cW$ to $\cW$, or
as an element of the Hilbert space $\otimes_\s^2\cW$. We will treat
the former interpretation as the standard one).

We will write
\[\gamma_\Phi:=\gamma_{\Phi,\Phi},\ \ \ \alpha_\Phi:=\alpha_{\Phi,\Phi}.\]
Note that
\beq  \alpha_{\Phi,\Psi}^\#=\alpha_{\Phi,\Psi},\eeq
\beq\left[\begin{array}{cc}\gamma_\Phi&\alpha_\Phi\\
   \bar \alpha_\Phi&\one+\bar{\gamma_\Phi}\end{array}\right]\geq0.\label{posi}\eeq
For further use note that (\ref{posi}) is equivalent to
\beq
\gamma_\Phi\geq0,\ \ \
\gamma_\Phi\geq\alpha_\Phi(\one+\bar{\gamma_\Phi})^{-1}\bar{\alpha_\Phi}.\eeq

Clearly, if $h$ is an operator on $\cW$ and $g\in\otimes_\s^2\cW$, then
\begin{eqnarray}
  (\Phi|\d\Gamma(h)\Psi)&=&\Tr\gamma_{\Phi,\Psi} h,\\
  (\Phi|\hat a^*(g)\Psi)&=&\Tr\alpha_{\Phi,\Psi}^*g,\label{ann1}\\
  (\Phi|\hat a(g)\Psi)&=&\Tr\alpha_{\Psi,\Phi}g^*.\label{ann2}
  \end{eqnarray}

Note that
(\ref{ann1}) and (\ref{ann2}) are still true if $g$ is an unbounded
functional on $\otimes_\s^2\cW$ with domain $\otimesal_\s^2\cD$,
provided that $\Psi,\Phi\in\Gammal_\s(\cD)$ where $\cD=\Dom h^{-\frac12}$,
as discussed in Remark \ref{form2}.

The following proposition provides a key estimate for the construction
of normally ordered Bogoliubov Hamiltonians:

\bep
Assume that $\|h^{-\frac12}g\bar h^{-\frac12}\|\leq 1$ and $\Tr g^*h^{-1}g<\infty$. Let
$\|h^{-\frac12}g\bar h^{-\frac12}\|\leq c$.
Then for $\Phi\in\Gamma_\s(\cW)$ with $\|\Phi\|=1$,
\beq(\Phi|\hat a^*(g)\Phi)\leq c(\Phi|\d\Gamma(h)\Phi)+\frac{1}{2c}\Tr(g^*h^{-1}g).\eeq
\label{iuy}\eep

\proof
\begin{eqnarray*}
  &&(\Phi|\hat a^*(g)\Phi)\ = \ |\Tr\bar{\alpha_\Phi }g|\\
    &=&\big|\Tr(\one+\bar{\gamma_\Phi})^{-\frac12}\bar\alpha_\Phi
  h^{\frac12}h^{-\frac12}g(\one+\bar{\gamma_\Phi})^{\frac12}\big|
  \\
&  \leq &\Big(\Tr h^{\frac12}\alpha_\Phi(\one+\bar{\gamma_\Phi})^{-1}\alpha_\Phi^*
  h^{\frac12}\Big)^{\frac12}
  \\&&\times\Big(\Tr
   h^{-\frac12}g(\one+\bar{\gamma_\Phi})g^*
   h^{-\frac12}\Big)^{\frac12}\\
   &\leq&
   \Big(\Tr h^{\frac12}\gamma_\Phi
  h^{\frac12}\Big)^{\frac12}\Big(\Tr
  h^{-\frac12}gg^*
  h^{-\frac12}+
\|h^{-\frac12}gh^{-\frac12}\|^2
\Tr h^{\frac12}
\bar{\gamma_\Phi}  h^{\frac12}\Big)^{\frac12}\\
&=&
   \Big((\Phi|\d\Gamma(h)\Phi)
  \Big)^{\frac12}\Big(\Tr
 g^* h^{-1}g
 +
\|h^{-\frac12}gh^{-\frac12}\|^2
(\Phi|\d\Gamma(h)\Phi)  \Big)^{\frac12}.
\end{eqnarray*}
Then we use the inequality
\beq \sqrt{x(y+c_0^2x)}\leq cx+\frac{y}{2c}\eeq
valid for $x,y\geq0$, $c>c_0$. \qed

\bet 
Assume that $\|h^{-\frac12}g\bar h^{-\frac12}\|< 1$ and $\Tr g^*h^{-1}g<\infty$.
Then the quadratic form
\beq
\d\Gamma(h)+\12\hat a^*(g)+\12\hat a(g)\label{posi1}\eeq
defined on the form domain of $\d\Gamma(h)$ is closed and bounded from below
by $-\12\Tr(g^*h^{-1}g)$. Hence it defines a self-adjoint operator,
which we temporarily denote by $C$. It satisfies
\beq\big(1+\d\Gamma(h)\big)^{\frac12}(\i +C)^{-1}
\big(1+\d\Gamma(h)\big)^{\frac12}\ \ \hbox{is bounded}.\label{pqo2}
\eeq
\label{th4}\eet

\proof By Proposition \ref{iuy},
\beq
\12|\big(\Phi|(\hat a^*(g)+\hat a(g))\Phi)|\leq c (\Phi|
\d\Gamma(h)\Phi)+\frac{1}{2c}\Tr g^*h^{-1}g\|\Phi\|^2.\eeq
Setting $c:=\|h^{-\frac12}g\bar h^{-\frac12}\|<1$ and using the KLMN
Theorem,
we see that the form 
(\ref{posi1}) is closed and bounded from below, and hence defines a
bounded from below self-adjoint operator $C$. Setting $c=1$, we see  that
\beq -\12\Tr(g^*h^{-1}g)<C.\eeq
(\ref{pqo2}) is also a  consequence of the KLMN Theorem. \qed

\bet The operator defined in Thm \ref{th4}
is the normally ordered quantization of $B$. In other words, following the notation introduced in Subsection
\ref{Bogoliubov Hamiltonians}, $C=\hat H_B^\n$.\label{thmw1}
\eet

On a formal level the above theorem is essentially obvious.  However,
there are technical difficulties, for which  we will need a few
technical
lemmas. In these lemmas we use $h\geq0$ and 
$\| h^{-\frac12}g\bar h^{-\frac12}\|<1$. Note that under this
assumption, $\i\tau$ belongs to the resolvent set of $B$ for  $\tau\neq0$. 

\bel For  $\tau\neq0$, $B(\tau^2+B^2)^{-1}$ has a dense range. \label{nor7}\eel

\proof
We write
\beq 
B(\tau^2+B^2)^{-1}=(\i\tau+B)^{-1}(-\i\tau+B)^{-1}B.\label{nor5}\eeq
We will show that 
(\ref{nor5}) has a dense range when restricted to $\Dom B$.

First note that $B=AS$ where $A$ is self-adjoint and $S$ is unitary.
Hence $\Dom B=S\Dom A$ and $B\Dom B=A\Dom A$. This shows that $B\Dom
B$ is dense.

Then we apply twice Lemma \ref{nor6} to the bounded operators with
dense range $(\i\tau+B)^{-1}$ and $(-\i\tau+B)^{-1}$.
\qed

\bel For  $\tau\neq0$, the operator
$A_0^{-\frac12}B(\tau^2+B^2)^{-1}$ is bounded.
\label{nor8}\eel

\proof
First note that
\beq
\big\|(\i \tau S+A_0)^{-\frac12}GS(\i \tau
S+A_0)^{-\frac12}\big\|=\big\|h^{\frac12}g \bar h^{\frac12}\big\|<1.\label{pqo1}\eeq
Next we check
that all the terms on the right of the following  identity are bounded:
\begin{eqnarray}
A_0^{-\frac12}B(\i\tau
+B)^{-1}&=&(1+A_0^{-\frac12}GSA_0^{-\frac12})\\&&\times
A_0^{\frac12}(\i \tau S+
A_0)^{-\frac12}\\&&\times
\big(1+(\i \tau S+A_0)^{-\frac12}GS(\i \tau S+A_0)^{-\frac12}\big)^{-1}
\label{pqo}
\\&&\times(\i \tau S+
A_0)^{-\frac12}.\end{eqnarray}
(To see that (\ref{pqo}) is well defined we use (\ref{pqo1})).
Therefore, $A_0^{-\frac12}B(\i\tau+B)^{-1}$ is bounded, which
obviously implies the boundedness of $A_0^{-\frac12}B(\tau^2+B^2)^{-1}$.
\qed

\noindent
{\bf Proof of Thm \ref{thmw1}}.
Consider $w\in\Ran B(B^2+1)^{-1}$. By Lemma \ref{nor7}, such $w$ are
dense in $\cW$.

 Set
\beq\left[\begin{array}{c}w_t\\\bar w_t\end{array}\right]:=
\e^{\i tB}\left[\begin{array}{c}w\\\bar w\end{array}\right].\eeq
By Lemma \ref{nor8}, $\|h^{-\frac12}w_t\|$ is uniformly bounded.
Therefore, by Prop. \ref{ntau},
\beq(1+\d\Gamma(h))^{-\frac12}\hat \phi(w_t,\bar
w_t)(1+\d\Gamma(h))^{-\frac12}\eeq
is uniformly bounded.
Hence, by (\ref{pqo2}), so is
\beq k(t):=(C+\i)^{-1}\e^{-\i tC}
\hat \phi(w_t,\bar w_t)
\e^{\i tC}(C+\i)^{-1}.\eeq

We know that $(w_t,\bar w_t)\in\Dom (B)$. But this does not
necessarily imply that
$w_t\in\Dom h$. It only implies $w_t\in\Dom h^{\frac12}$. Therefore,
strictly speaking,
we cannot write
\beq\frac{\d}{\d t}w_t=\i hw_t-\i g \bar
w_t,\eeq
but only
\beq\frac{\d}{\d t}h^{-\frac12}w_t=\i h^{-\frac12}hw_t-\i h^{-\frac12}g \bar w_t.\eeq
However, using the boundedness of $(\i
+C)^{-1}\big(\i+\d\Gamma(h)\big)^{\frac12}$ and Prop.  \ref{ntau},
it is sufficient to compute:
\begin{eqnarray}&&
 (\i+C)^{-1} \i\big[C,\hat\phi(w_t,\bar w_t)\big](\i+C)^{-1}\\
  &=&(\i+C)^{-1}\Big(
  \hat a^*(h w_t)+\hat a(g \bar w_t)-\hat a(h w_t)-\hat a^*(g\bar
  w_t)\Big)(\i+C)^{-1}\\
  &=&(\i+C)^{-1}\hat \phi\big(\tfrac{\d}{\d t} w_t,\tfrac{\d}{\d t} \bar
  w_t\big)(\i+C)^{-1}.\end{eqnarray}
Therefore,
\begin{eqnarray}
  \frac{\d}{\d t}k(t)&=&
(C+\i)^{-1}\e^{-\i tC}
\Big(-\i\big[C,\big(\hat \phi(w_t,\bar w_t)\big)\big]\\
&&\ \ +\hat \phi\big(\tfrac{\d}{\d t}w_t,\tfrac{\d}{\d t}\bar w_t)\Big)
\e^{\i tC}(C+\i)^{-1}\ =\ 0.\end{eqnarray}
 This shows that $k(t)$ does not
  depend on $t$. Therefore,
  \beq
  (C+\i)^{-1}\e^{\i tC}
\hat \phi(w,\bar w)
\e^{-\i tC}(C+\i)^{-1}
=(C+\i)^{-1}
\hat\phi(w_t,\bar w_t)(C+\i)^{-1}.\eeq
This proves that $\e^{\i tC}$ implements $\e^{\i tB}$.

Clearly, $\gamma_\Omega=0$, $\alpha_\Omega=0$, and $\Omega\in
\Dom(\d\Gamma(h)^{\frac12})=\Dom(|C|^{\frac12})$. Therefore,
\beq (\Omega|C\Omega)=0.
\eeq
Thus, by Proposition \ref{normal},
the operator temporarily denoted
$C$ is the normally ordered quantization of $B$.
\qed

\subsection{Infimum of  normally ordered Hamiltonians}
\label{Infimum of  normally ordered Hamiltonians II}

In Subsection
\ref{Infimum of normally ordered Hamiltonians}, in the finite
dimensional context,
we defined $E_B^\n$ as
the infimum of the normally ordered Hamiltonian $\hat H_B^\n$.
In infinite dimension it is useful to define $E_B^\n$ independently of
whether $\hat H_B^\n$ exists or not.

As a basic condition on the symplectic generator $B$ we
assume that $h>0$,  $\|h^{-\frac12}g\bar h^{-\frac12}\|< 1$.
As in (\ref{fa1}), for $\sigma\in\rr$ we set
\beq A_\sigma:=A_0+\sigma GS=
\left[\begin{array}{cc}h&\sigma
      g\\\sigma \bar g&\bar
      h\end{array}\right],\eeq
so that $A=A_1$.

Out of the formulas for $E^\n$ listed in (\ref{fs1})--(\ref{fs5})
valid in finite dimensions, the most suitable for infinite dimensions
seems to be (\ref{fs4a}), which we choose as the definition of $E^\n$:
 \beq E^\n_B:=\frac18\int_0^1\d\sigma\Tr
 A_\sigma^{\frac12}(A_\sigma^{\frac12}SA_\sigma
SA_\sigma^{\frac12})^{-\frac12}
 A_\sigma^{\frac12}GS,\label{fs5c}\eeq
provided that the above integral is well defined.

(\ref{fs5}) is another formula for $E^\n$ useful in
infinite dimension:
\bep We have
\begin{eqnarray}
E^\n_B  &:=&\frac{1}{8}\int_0^1\d\sigma\int\frac{\d\tau}{2\pi}(1-\sigma)
\Tr\frac{1}{(A_\sigma+\i\tau S)}SG\frac{1}{(A_\sigma+\i\tau S)}SG. \label{fs5a-}
\end{eqnarray}
More precisely, if (\ref{fs5a-})
is well defined as a convergent integral, then it coincides with
(\ref{fs5c}).\eep

Below we list a few criteria for the existence of $E^\n_B$.
\bet 
\ben\item Let $\|g\|_1<\infty$. Then $E^\n_B$ is well defined by
(\ref{fs5c}).
\item Let $s_-<\frac12<s_+$. Suppose that
and $\Tr
g\bar h^{-s_-}g^* h^{-s_-}<\infty$ and
 $\Tr
g\bar h^{-s_+}g^* h^{-s_+}<\infty$.  
 Then $E^\n_B$ is well defined by
 (\ref{fs5c}) or (\ref{fs5a-}).
\item
Suppose that  $\Tr
gh^{-1}g^*<\infty$.   Then $E^\n_B$ is well defined by
(\ref{fs5c}) or (\ref{fs5a-}).
 \een
\label{th5}\eet

\proof (1): We apply Inequality (\ref{ineq0})
to the operator $Y:= A_\sigma^{\frac12}(A_\sigma^{\frac12}SA_\sigma
SA_\sigma^{\frac12})^{-\frac12}
A_\sigma^{\frac12}$ 
and  $X:=GS$. Note that $Y$ is uniformly bounded
for $\sigma\in[0,1]$. We obtain
\begin{eqnarray}
  | E^\n|&\leq&\frac18\int_0^1\d\sigma\big|\Tr
 A_\sigma^{\frac12}(A_\sigma^{\frac12}SA_\sigma
SA_\sigma^{\frac12})^{-\frac12}
 A_\sigma^{\frac12}GS\big|\\
&\leq&\frac18\int_0^1\d\sigma
\big\| A_\sigma^{\frac12}(A_\sigma^{\frac12}SA_\sigma
SA_\sigma^{\frac12})^{-\frac12}
 A_\sigma^{\frac12}\big\|\Tr\sqrt{G^2}\leq c\Tr\sqrt{G^2}.\end{eqnarray}
But $\Tr\sqrt{G^2}=2\Tr \sqrt{\bar gg}$. This proves (1).

 (2):
First note that
 for $0\leq s\leq 1$
\beq \|A_0^{\frac{s}{2}}(A_0+\i\tau S)^{-\frac12}\|\leq \tau^{-\frac12+\frac{s}2}
.\label{ineqa1}
\eeq
Indeed,
\beq
A_0^{\frac{s}{2}}(A_0+\i\tau S)^{-\frac12}=
A_0^{\frac{s}{2}}(A_0+\i\tau S)^{-\frac{s}{2}}\times(A_0+\i\tau S)^{-\frac12+\frac{s}{2}},
\eeq
where  the first term
is bounded by $1$ and the second is bounded by $\tau^{-\frac12+\frac{s}2}$.

Moreover,
\beq
\|(A_0+\i\tau S)^{-\frac12}GS(A_0+\i\tau S)^{-\frac12}\|\leq
\|A_0^{-\frac12}GSA_0^{-\frac12}\|=a<1.
\label{ineqa2}
\eeq
Therefore, we can write
\beq
(A_\sigma+\i\tau S)^{-1}
=(A_0+\i\tau S)^{-\frac12}
\big(\one-\sigma(A_0+\i\tau S)^{-\frac12}GS(A_0+\i\tau
S)^{-\frac12}\big)^{-1}(A_0+\i\tau S)^{-\frac12}.
\label{ineq1}\eeq

(\ref{ineq1}) together with (\ref{ineqa1}) and (\ref{ineqa2})
yield
\beq
\big\|A_0^{\frac{s}{2}}(A_\sigma+\i\tau S)^{-1}A_0^{\frac{s}{2}}\big\|
\leq (1-a)^{-1}\tau^{-1+s}.\label{ineq4}\eeq

Now,
\begin{eqnarray}
  &&\Big|\Tr\frac{1}{(A_\sigma+\i\tau S)}GS\frac{1}{(A_\sigma+\i\tau
    S)}G\Big|\\
  &\leq&
  \Big\|A_0^{\frac{s}{2}}(A_\sigma+\i\tau S)^{-1}A_0^{\frac{s}{2}}
  \Big\|^2\Tr GA_0^{-s}G^*A_0^{-s},\label{ineq2}\\
  &\leq&(1-a)^{-2}\tau^{2s-2}\Tr GA_0^{-s}G^*A_0^{-s},
  \end{eqnarray}
where we first used Inequality (\ref{ineq}) with
$Y=Z:=A_0^{\frac{s}{2}}(A_\sigma+\i\tau S)^{-1}A_0^{\frac{s}{2}}$ and $X:=
A_0^{-\frac{s}{2}}G A_0^{-\frac{s}{2}} $, and then we applied
(\ref{ineq4}). Thus
\begin{eqnarray}
  |E^\n|
 &\leq&\frac{1}{8}\int_0^1\d\sigma\int\frac{\d\tau}{2\pi}(1-\sigma)
\Big|\Tr
\frac{1}{(A_\sigma+\i\tau S)}GS\frac{1}{(A_\sigma+\i\tau S)}G\Big|\\
  &\leq&
c\Tr GA_0^{-s_+}G^*A_0^{-s_+}\int_0^1\tau^{2s_+-2}\d \tau+c\Tr GA_0^{-s_-}G^*A_0^{-s_-}
\int_1^\infty \tau^{2s_--2}\d \tau.
\end{eqnarray}
But $\Tr GA_0^{-s_\pm}GA_0^{-s_\pm}=2\Tr\bar g h^{-s_\pm} g\bar h^{-s_\pm}$.
 This proves (2).

(3):
Applying (\ref{ineq}) to
\[Y=Z:=\big(\one-\sigma(A_0+\i\tau S)^{-\frac12}GS(A_0+\i\tau
S)^{-\frac12}\big)^{-1},\ \ \ X:=
(A_0+\i\tau S)^{-\frac12}GS(A_0+\i\tau S)^{-\frac12},\]
and using (\ref{ineq1}), (\ref{ineqa2}), we obtain
\begin{eqnarray}
&&\Big|\Tr\frac{1}{(A_\sigma+\i\tau S)}GS\frac{1}{(A_\sigma+\i\tau S)}GS\Big|
  \\
&=&\Tr XYXY\leq\|Y\|^2\Tr XX^*\\
&\leq&
\frac{1}{(1-\sigma a)^2}\Tr\frac{1}{(A_0^2+\tau^2)^{\frac12}}G
\frac{1}{(A_0^2+\tau^2)^{\frac12}}G^*\\
&\leq&\frac{1}{(1-\sigma a)^2}\Tr G\frac{1}{(A_0^2+\tau^2)}G^*.
\end{eqnarray}
Therefore,
\begin{eqnarray}
  |E^\n|
 &\leq&\frac{1}{8}\int_0^1\d\sigma\int\frac{\d\tau}{2\pi}(1-\sigma)
\Big|\Tr
\frac{1}{(A_\sigma+\i\tau S)}GS\frac{1}{(A_\sigma+\i\tau S)}GS\Big|\\
  &\leq&
  \frac{1}{8}\int_0^1\d\sigma\frac{(1-\sigma)}{(1-a\sigma)^2}
  \int\frac{\d\tau}{2\pi}
  \Tr   G\frac{1}{(A_0^2+\tau^2)}G^*
  \\&=&
  \frac{(-\log(1-a)-a)}{8a^2}\Tr G\frac{1}{A_0}G^*.\end{eqnarray}
But $\Tr G\frac{1}{A_0}G=2\Tr gh^{-1}g^*$. 
This proves (3)
\qed

\bet 
  Suppose that  $\Tr
gh^{-1}g^*<\infty$,
  as in Thm \ref{th5} (3). Let  $\hat H^\n_B$ be defined
  as in Subsect. \ref{Normally ordered Hamiltonian}.
Let $E^\n_B$ be defined as in (\ref{fs5c}).
  Then
  \beq E^\n_B=\inf \hat H^\n_B.\label{fs10}\eeq
  \label{nor3}
  \eet

If $\cW$ is finite dimensional, then (\ref{fs10}) was proven in
Thm \ref{thm-nor}. In our proof, we will reduce the full problem to
this case.
  The proof will be divided into several steps.

  \noindent{\bf Step 1.} {\em Suppose that there exists a finite
  dimensional $\cW_0$ such that $\Ran g\subset\cW_0$ and $h$ preserves
  $\cW_0$. Then (\ref{fs10}) is true.}

\proof Set $\cW_1:=\cW_0^\perp$. Note that $g=g^\t$ implies that
$\bar\cW_1\subset\Ker g$. Let $h_0,g_0$ denote the restrictions of
$g,h$ to $\cW_0$. Let $h_1$ denotes the restriction of $h$ to $\cW_1$.
Consider the symplectic generator
 on $\cW_{0}$ \beq B_{0}:=\left[\begin{array}{cc}h_{0}&
      -g_{0}\\ \bar g_{0}&
     -\bar h_{0}\end{array}\right],\eeq
and the corresponding normally ordered Bogoliubov Hamiltonian
\beq \hat
H_0^\n:=\hat H_{B_{0}}^\n=\d\Gamma(h_{0})+\frac12\big(\hat a^*(g_{0})+\hat a(g_{0})\big)
.\eeq 
We will write  $E_{0}^\n$, resp. $E^\n$ for
 $E_{B_0}^\n$, resp. $E_B^\n$.

We have the decomposition
\beq\Gamma_\s(\cW)\simeq\Gamma_\s(\cW_{0})\otimes\Gamma_\s(\cW_{1})
\eeq
The operator $\hat H^\n$ can be decomposed as
\beq
\hat H^\n\simeq
\hat
H_{0}^\n\otimes\one
+\one\otimes\d\Gamma(h_{1}).\eeq
We have
\begin{eqnarray}\inf\hat H^\n&=&
\inf\hat
H_{0}^\n=E_0^\n=E^\n,\label{nor2}\end{eqnarray}
where in the middle step we used the finite dimension of $\cW_{0}$.
 \qed

 \noindent{\bf Step 2.} {\em Suppose that $g$ is finite dimensional
   and $\one_{[\delta,\delta^{-1}]}(h)g=g$. Then (\ref{fs10}) is true.}

   \proof
Let $\epsilon>0$.   Let us set
\begin{eqnarray}
  \pi_{\epsilon,n}&:=&\one_{[(1+\epsilon)^n,(1+\epsilon)^{n+1}[}(h),\\
      h_\epsilon&:=&\sum_{n=-\infty}^\infty(1+\epsilon)^{n+1}\pi_{\epsilon,n}.
\end{eqnarray} 
    Note that
    \beq (1+\epsilon)^{-1} h_\epsilon\leq h\leq h_\epsilon.\eeq
    Hence,
    \beq  \d\Gamma\big((1+\epsilon)^{-1}h_\epsilon\big)\leq\d\Gamma(h)
    \leq\d\Gamma(h_\epsilon).\eeq
    Now
    \begin{eqnarray*}
\hat H_{\epsilon,-}^\n:=      \d\Gamma\big((1+\epsilon)^{-1}h_\epsilon\big)+\frac12\big(\hat
a^*(g)+\hat a(g)\big)&\leq\hat H^\n\leq&
 \d\Gamma\big(h_\epsilon\big)+\frac12\big(\hat
 a^*(g)+\hat a(g)\big)=:H_{\epsilon,+}^\n.
 \end{eqnarray*}
        
Let $\cW_{\epsilon,0}$ be the smallest subspace of $\cW$ containing
$\Ran g$ and left invariant by $h_\epsilon$.
 In other words,
\beq\cW_{\epsilon,0}:=\Span\{\pi_{\epsilon,n} w\ :\ w\in\Ran
g\}.\eeq
Note that  $\pi_{\epsilon,n}\Ran g=0$ for $|n|$ large
enough.  Therefore,
$\cW_{\epsilon,0}$ is finite dimensional.

Thus $\hat H_{\epsilon,\pm}^\n$ satisfy the conditions of Step 1, and so
\beq \inf \hat H_{\epsilon,\pm}^\n=E_{\epsilon,\pm}^\n,\eeq
in the obvious notation.
Using Lemma \ref{iuy1} we show that
\beq \lim_{\epsilon\to0}E_{\epsilon,\pm}^\n=E^\n,\eeq

Besides,
\beq \inf\hat H_{\epsilon,-}^\n\leq\inf H^\n\leq
\inf \hat H_{\epsilon,+}^\n.\eeq
\qed

 \noindent{\bf Step 3.} {\em Suppose that 
   for some $\delta>0$ we have
   $\one_{[\delta,\delta^{-1}]}(h)g=g$. Then (\ref{fs10}) is true.}

 \proof
 We know that $h^{-\12}g$ is Hilbert-Schmidt. Finite dimensional
 operators are dense in Hilbert-Schmidt operators. Therefore,
 given $\epsilon>0$, we can find a finite dimensional
$g_\epsilon$ such that $g_\epsilon=g_\epsilon^\t$,
$\one_{[\delta,\delta^{-1}]}(h)g=g$ and
\beq\big\|h^{-\12}g-h^{-\12}g_\epsilon\big\|_\HS=
\sqrt{\Tr(\bar g-\bar g_\delta')h^{-1}( g-
  g_\delta')}<\epsilon^2.\label{combi3}\eeq
Now, the Hilbert-Schmidt norm dominates the operator norm. Hence,
(\ref{combi3}) implies
\beq\|h^{-\frac12}(g-g_\epsilon)\|\leq\epsilon.\eeq
As a consequence,
\beq\|h^{-\frac12}(g-g_\epsilon)\bar
h^{-\frac12}\|\leq\epsilon\delta^{-\frac12}.\label{combi2}\eeq 

Therefore, by (\ref{combi2}),
\begin{eqnarray}
  \|h^{-\frac12}g_\epsilon \bar h^{-\frac12}\|&\leq &\|h^{-\frac12}g \bar
h^{-\frac12}\|+
\|h^{-\frac12}(g -g_\epsilon)\bar h^{-\frac12}\|\\
&\leq& a+\epsilon\delta^{-\frac12}=:a_1.\end{eqnarray}
By choosing $\epsilon$
small enough we can guarantee that $a_1<1$.

Now
\begin{eqnarray}
  \hat H^\n&=&(1-\nu)\d\Gamma(h)+\frac12\big(\hat
  a^*(g_\epsilon)+\hat
  a(g_\epsilon)\big)\\&&+\nu\d\Gamma(h)+\frac12\big(\hat
  a^*(g-g_\epsilon)+\hat a(g-g_\epsilon)\big)\\ 
&\geq&(1-\nu)\d\Gamma(h)+\frac12\big(\hat
  a^*(g_\epsilon)+\hat
  a(g_\epsilon)\big)-\frac{\epsilon^2}{\nu},\\
  \hat H^\n&=&(1+\nu)\d\Gamma(h)+\frac12\big(\hat
  a^*(g_\epsilon)+\hat
  a(g_\epsilon)\big)\\&&-\nu\d\Gamma(h)+\frac12\big(\hat
  a^*(g-g_\epsilon)+\hat a(g-g_\epsilon)\big)\\ 
&\leq&(1+\nu)\d\Gamma(h)+\frac12\big(\hat
  a^*(g_\epsilon)+\hat
  a(g_\epsilon)\big)+\frac{\epsilon^2}{\nu}.
  \end{eqnarray}
The Hamiltonians 
$(1\pm\nu)\d\Gamma(h)+\frac12\big(\hat
  a^*(g_\epsilon)+\hat
  a(g_\epsilon)\big)$ satisfy the assumptions of Step 2. \qed

 \noindent{\bf Step 4.} {\em  (\ref{fs10}) is true without additional
   assumptions.} 

\proof
 $h^{-\frac12}g$ is Hilbert-Schmidt
and $\slim\limits_{\delta\to0}\one_{[\delta,\delta^{-1}]}(h)=\one$. Hence, for any
$\epsilon>0$  we can
find $1\geq \delta>0$ such that if we set
\beq
g_\delta:=\one_{[\delta,\delta^{-1}]}(h)g\one_{[\delta,\delta^{-1}]}(\bar
h),\eeq
then
\beq\big\|h^{-\12}g-h^{-\12}g_\delta'\big\|_\HS=
\sqrt{\Tr(\bar g-\bar g_\delta)h^{-1}( g-
  g_\delta)}<\epsilon.\eeq
Note that
\beq
\|h^{-\frac12}g_\delta \bar
h^{-\frac12}\|\leq \|h^{-\frac12}g \bar
h^{-\frac12}\|=a.\label{combi1}\eeq
Then we estimate similarly as at the end of Step 3. We argue, that we
need only estimates about the Hamiltonians
$(1\pm\nu)\d\Gamma(h)+\frac12\big(\hat
  a^*(g_\delta)+\hat
  a(g_\delta)\big)$, which satisfy the assumptions of Step 3. \qed

A result about the continuity of $E_B^\n$ with respect to $h$, which we needed in the above proof, is described below.

  \bel Let 
\beq
B=\left[\begin{array}{cc}h&-g\\\bar g&-\bar h\end{array}\right],
\ \ \ B'=\left[\begin{array}{cc}h'&-g\\\bar g&-\bar h'\end{array}\right].
\label{gene1}\eeq
with
\[
\|h^{-\frac12}g \bar
h^{-\frac12}\|,\ \|h^{\prime-\frac12}g \bar
h^{\prime-\frac12}\|\leq a,\ \ \ \ \ 
\Tr gh^{-1}g^*,\ \Tr gh^{\prime-1}g^*\leq a_1
\]
Then
\beq\big\|E_B^\n-E_{B'}^\n\big\|\leq c
\|h^{-\frac12}(h-h') 
h^{\prime-\frac12}\|,
\eeq
where $c$ depends only on $a$ and $a_1$.
\label{iuy1}
\eel

\proof
\begin{eqnarray*}&&
E^\n_B -E_{B'}^\n\\ &=&\frac{1}{8}\int_0^1\d\sigma\int\frac{\d\tau}{2\pi}(1-\sigma)
\Tr\frac{1}{(A_\sigma+\i\tau S)}(A_0'-A_0)\frac{1}{(A_\sigma'+\i\tau S)}
SG\frac{1}{(A_\sigma+\i\tau S)}SG\\
 &&\!\!\!\!\!\!+\frac{1}{8}\int_0^1\d\sigma\int\frac{\d\tau}{2\pi}(1-\sigma)
\Tr\frac{1}{(A_\sigma'+\i\tau S)}SG\frac{1}{(A_\sigma+\i\tau S)}(A_0'-A_0)
\frac{1}{(A_\sigma'+\i\tau S)} 
SG.
\end{eqnarray*}
Then we argue similarly as in the proof of Theorem
\ref{th5} (3). \qed

\subsection{Weyl Bogoliubov Hamiltonian}
\label{Weyl Bogoliubov Hamiltonian}

Weyl Bogoliubov Hamiltonians play a central role in the
theory of Bogoliubov Hamiltonians, providing the simplest algebraic
formulas.
Unfortunately, in infinite
dimensions they are usually
ill-defined.

If $A_B\geq0$, then we can define
\beq E_B^\w:
= \frac{1}{4}\Tr
\sqrt{A_B^{\frac12}SA_BSA_B^{\frac12}},
\label{functa}\eeq
which is a nonnegative number, often infinite.
Recall that in finite dimension it coincides with the infimum of $\hat
H_B^\w$.

The following theorem gives (rather restrictive)
conditions when we can define the Weyl quantization in any dimension.

\bet 
Assume that $h>0$,  $\|h^{-\frac12}g\bar h^{-\frac12}\|=:a< 1$ and $\Tr
h<\infty$.  Then the following holds:
\ben \item
$  \Tr g^*h^{-1}g<\infty.$\label{func4}
\item  $\|g\|_1<\infty.$\label{func4a}
\item By (1), we can define  $\hat H_B^\n$ as in Subsect. \ref{Normally ordered
  Hamiltonian},
$E_B^\n$ is well defined as in Subsect
\ref{Infimum of  normally ordered Hamiltonians II}, and by Thm  
\ref{nor3},
\beq \inf H_B^\n= E_B^\n.\label{pq1}\eeq

\item By  Thm
\ref{th2} (3),  $H_B^\w$ is well defined. We have
\beq \hat H_B^\w:=\hat H_B^\n+\12\Tr h.\label{tru}\eeq

\item
$\sqrt{A_B^{\frac12}SA_BSA_B^{\frac12}}$  is trace class, so that
$E_B^\w$ is finite. We have
\beq E_B^\w=E_B^\n+\frac12\Tr h\label{pq2}\eeq
\beq E_B^\w
=\inf \hat H_B^\w.\label{func5}\eeq\een
\label{th3}\eet

\proof
We have
\beq
h^{-\frac12}gg^*h^{-\frac12}=h^{-\frac12}g\bar h^{-\frac12}\bar h
\bar h^{-\frac12}g^*h^{-\frac12}.
\label{tra}\eeq
But $h^{-\frac12}g\bar h^{-\frac12}$ and
$\bar h^{-\frac12}g^*h^{-\frac12}$ are bounded and $\bar h$ is trace
class. Therefore,
(\ref{tra}) is trace class. Hence (1) is true.

\beq
\|g\|_1=\|h^{\frac12}h^{-\frac12}g\bar h^{-\frac12}\bar
h^{\frac12}\|_1
\leq
\|h^{\frac12}\|_2\|h^{-\frac12}g\bar h^{-\frac12}\|\|\bar
h^{\frac12}\|_2\leq a\|h\|_1
\eeq
proves (2).

We know that $h$ is trace class. Besides, $\|g\|_1<\infty$ implies
$\|g\|_\HS<\infty$. Therefore, the assumptions of
Thm \ref{th2} (3) are satisfied. Therefore, $\hat H_B^\w$ is well defined.

By (\ref{hamub}), (\ref{hamub0})
\beq 
\sqrt{A^{\frac12}SASA^{\frac12}}\leq\frac{(1+a)^{\frac12}}{(1-a)^{\frac12}}A\leq
\frac{(1+a)^{\frac32}}{(1-a)^{\frac12}}A_0.\label{tra1}
\eeq
But $A_0$ is trace class. Hence so is the left hand side of
(\ref{tra1}).

By
repeating the arguments of Thm \ref{thm-nor}
we see that (\ref{pq2}) is true.
By Thm \ref{th2} (3) we have (\ref{tru}).
Combining (\ref{pq1}), (\ref{pq2}) and (\ref{tru}), we obtain
(\ref{func5}).
\qed

\subsection{Infimum of the renormalized Hamiltonian}

Recall that in Subsections
\ref{Renormalization I} and \ref{Renormalization II} we discussed
the renormalized Hamiltonians $\hat H_B^\ren$ and its infimum
$E_B^\ren$ in the context of finite dimension. These objects are of
course especially interesting in infinite dimensions.

Note that it may happen
that $E_B^\ren$ is well defined and $\hat H_B^\ren$ is not. In this
subsection we discuss only
 $E_B^\ren$, without asking whether  $\hat H_B^\ren$ exists.

We will use the framework of Subsection
\ref{Renormalization II} with $\lambda=1$.
That means, we assume that $h_0>0$, $h=h_0+h_1$ and $ h_1^2=g\bar
g$, $h_1g=g\bar h_1$.  Recall that
we have
\begin{eqnarray}
B_0&=&\left[\begin{array}{cc}h_0&0\\0&-\bar
    h_0\end{array}\right],\\
  A_0=B_0S&=&\left[\begin{array}{cc}h_0&0\\0&\bar h_0\end{array}\right],\\
  B_0^2=A_0^2&=&\left[\begin{array}{cc}h_0^2&0\\0&\bar
      h_0^2\end{array}\right],\label{fa1c}\\
V=B^2-B_0^2&=&\left[\begin{array}{cc}h_0h_1+h_1h_0&-h_0g+g\bar
    h_0\\\bar gh_0-\bar h_0
  \bar  g&\bar h_0\bar h_1+\bar h_0\bar h_1
  \end{array}\right].\label{fa2a-}
\end{eqnarray}

We use (\ref{eren}) to define $E_B^\ren$:
\begin{eqnarray}
E^\ren&:=&-\frac14\int\Tr\frac{1}{B_0^2+\tau^2}V\frac{1}{B^2+\tau^2}\Big(V\frac{1}{B_0^2+\tau^2}\Big)^2
\tau^2\frac{\d\tau}{2\pi}
,\end{eqnarray}
provided that the above integral is well defined.
Below we give a simple criterion for the well definedness of $E^\ren$.

\bet
Suppose that
\beq \big\| B_0^{-1}V
B_0^{-1}\big\|=:a_1< 1,\label{pqs}\eeq and 
\begin{eqnarray}
L_3&:=&\frac{1}{4\cdot6}\int\Tr\Big(V\frac{1}{B_0^2+\tau^2}\Big)^3
\frac{\d\tau}{2\pi},\\
L_4&:=&-\frac{1}{4\cdot8}\int\Tr\Big(V\frac{1}{B_0^2+\tau^2}\Big)^4
\frac{\d\tau}{2\pi}
\end{eqnarray}
are finite. (In the case of $L_4$ the meaning of the assumption is
clear, since the integrand is always positive. This does not need to
be the case of $L_3$---here we assume that the integrand is integrable).  Then $E^\ren$ is well defined.
\label{pio}\eet

\proof
Assumption (\ref{pqs}) is equivalent to
\beq a_1  B_0^2\leq V\leq a_1 B_0^2.\eeq
Therefore,
\beq B^2=B_0^2+V\geq (1-a_1)B_0^2.\eeq
Hence, with $c:=\frac{1}{1-a_1}>0$,
\beq\frac{1}{(B^2+\tau^2)}\leq c\frac{1}{(B_0^2+\tau^2)}.\eeq
Besides,
\begin{eqnarray}
  E^\ren&:=&\frac{1}{4\cdot6}\int\Tr\Big(V\frac{1}{B_0^2+\tau^2}\Big)^3
\frac{\d\tau}{2\pi}\\
&&-\frac14\int\Tr\Big(\frac{1}{B_0^2+\tau^2}V\Big)^2
\frac{1}{B^2+\tau^2}\Big(V\frac{1}{B_0^2+\tau^2}\Big)^2
\tau^2\frac{\d\tau}{2\pi}
.\end{eqnarray}
The first term is precisely $L_3$. The second term is controlled by
$L_4$, because
\begin{eqnarray}
0  &\leq&
\frac14\int\Tr\Big(\frac{1}{B_0^2+\tau^2}V\Big)^2
\frac{1}{B^2+\tau^2}\Big(V\frac{1}{B_0^2+\tau^2}\Big)^2
\tau^2\frac{\d\tau}{2\pi}\\
&\leq &\frac{c}4\int\Tr
\Big(\frac{1}{B_0^2+\tau^2}V\Big)^2
\frac{1}{B_0^2+\tau^2}\Big(V\frac{1}{B_0^2+\tau^2}\Big)^2
\tau^2\frac{\d\tau}{2\pi}\label{cyc1}\\
&=&\frac{c}{4\cdot8}\int\Tr\Big(V\frac{1}{B_0^2+\tau^2}\Big)^4
\frac{\d\tau}{2\pi}\ =\ -cL_4.\label{cyc2}
\end{eqnarray}
To pass from (\ref{cyc1}) to
(\ref{cyc2}) we use identity  (\ref{cyclic}), which involves
 integration by parts.
\qed
\appendix
\section{Appendix}

\subsection{Complex conjugate space}
\label{Complex conjugate space}
\init
This appendix
is a side remark about complex conjugate spaces. This well known but
abstract and somewhat confusing concept appears naturally in the
context of Bogoliubov Hamiltonians. We follow  \cite{DG}.

Let $\cW$ be a Hilbert space. By definition, a space {\em complex conjugate
to} $\cW$ is another Hilbert space $\bar\cW$ equipped with a fixed
anti-unitary map $\chi:\cW\to\bar\cW$.

In the literature, various authors use
several concrete realizations of $\chi$ and
$\bar\cW$.

\ben\item We can assume that $\cW=\bar\cW$ and $\chi$ is
antiunitary on $\cW$ satisfying $\chi^2=\one$. Suppose that we choose a
basis fixed by $\chi$ (which is always possible).
Then $\chi$ amounts to conjugating the
components of vectors in this basis.
\item We can also identify $\bar\cW$ with
  the space of continuous linear functionals on
  $\cW$. We then define $\chi$ to be the {\em Riesz isomorphism}, that is,
  \beq\langle \chi z|w\rangle:=(z|w),\eeq
  (where $(\cdot|\cdot)$ denotes the scalar product and
  $\langle\cdot|\cdot\rangle$ the action of a linear functional). If we
  choose an orthonormal basis in $\cW$ and the dual basis in
  $\bar\cW$, then again $\chi$ amounts to conjugating the
components of vectors in this basis.
    \item 
Finally, one can set $\bar\cW=\cW$ as the real vector space, changing
only the complex structure to the opposite one and the scalar product
to the complex conjugate of the original scalar product. $\chi$ is defined to
be the identity operator.
If we fix any basis,
then similarly as before,
 $\chi$ is conjugating the components of vectors.
\een

The interpretation (1) is the most naive one. It is often
natural--especially if $\cW$ is 
$L^2$ of some measure space. It is used eg. in Sect. 1 or
in \cite{HS}. The interpretation (2) is used in \cite{NNS}. The
interpretation (3) is probably the most ``orthodox'' option--it does
not invoke anything besides the vector space
structure.
In particular, it does not involve the Hilbert space
structure of $\cW$.

Note that for all three interpretations, in typical bases, the action of
$\chi$ is equivalent to conjugating components of vectors. Similarly
 $\chi p\chi^{-1}$ and $\chi q\chi$ amounts to conjugating matrix
elements of $p$ and $q$.

\subsection{$*$-automorphisms of the algebra of bounded operators}

Let $\cH$ be a Hilbert space. A bijective linear map $\alpha$ on $B(\cH)$ is a
{\em $*$-automorphism} if
\beq\alpha(BC)=\alpha(B)\alpha(C),\ \ \alpha(C^*)=\alpha(C)^*,\ \ B,C\in
B(\cH).\eeq 

\bep (Example 3.2.14, \cite{BR1}). The following statements are
equivalent:
\ben\item
$\alpha$ is a  $*$-automorphism of $B(\cH)$.
\item There exists a unitary $U\in B(\cH)$ such that
\beq  \alpha(C)=UCU^*,\ \ C\in B(\cH),\eeq
\een
If (1),(2) hold, then  $U$ is determined uniquely up to a
phase factor.\label{auto1} \eep

Let $\rr\ni t\mapsto\alpha_t$ be a 1-parameter group of $*$-automorphisms of
$B(\cH)$. We say that  it is a {\em $C_0^*$-group} if
$t\mapsto\alpha_t(C)$ for any $C\in B(\cH)$ is weakly continuous.

\bep (Example 3.2.35, \cite{BR1}).
 The following statements are
equivalent:
\ben\item
$t\mapsto \alpha_t$ is  a $C_0^*$-group of
$*$-automorphisms  of
$B(\cH)$.
\item There exists self-adjoint $H$ on $\cH$ such that
  \beq \alpha_t(C)=\e^{\i tH}C\e^{-\i tH},\ \ C\in B(\cH).\eeq
  \een
If (1), (2) hold, then  $H$ is defined uniquely
up to an additive real constant. \label{auto2}\eep

\subsection{Useful identities and inequalities}
\beq
\int\frac{\d\tau}{2\pi}\frac{1}{(C^2+\tau^2)}=\frac{1}{\sqrt{C^2}}.\label{iden1}\eeq
\beq
\int\frac{\tau^2\d\tau}{2\pi}\frac{1}{(C^2+\tau^2)^2}=\frac{1}{4\sqrt{C^2}}.\label{app2}\eeq
\beq |\Tr XY|\leq\|Y\|\Tr\sqrt{XX^*}.\label{ineq0}\eeq
\beq
|\Tr XYXZ|\leq\sqrt{\Tr XYY^*X^*}\sqrt{X^*Z^*ZX}\leq\|Y\|\|Z\|\Tr XX^*.\label{ineq}\eeq
\beq\|a_{11}\|_1+\|a_{22}\|_1\leq \left\|\left[\begin{array}{cc}a_{11}&a_{12}\\
    a_{21}&a_{22}\end{array}\right]\right\|_1.
\label{traceclass}
\eeq

\subsection{Useful lemmas}

\bel Let C be a bounded operator on $\cW$ with a dense range. Let
$\cD$ be a dense subspace of $\cW$. Then $C\cD$ is dense.
\label{nor6}\eel

\proof Let $w\in\cW$. We can find $v_1\in\cW$ such that
$\|w-Cv_1\|<\frac{\epsilon}{2}$. We can find $v_2\in\cD$ such that
$\|v_1-v_2\|<\frac{\epsilon}{2\|C\|}$. Now
\beq\|w-Cv_2\|<\|w-w_1\|+\|Cv_1-Cv_2\|<\epsilon.
\eeq\qed

\bel Suppose that $h_1,h_2$ are self-adjoint operators on $\cW$ such
that for any $w,w'\in\cW$
\beq\lim_{t\to0}\frac1t\Big(\big(w|\e^{\i th_1}w'\big)-\big(w|\e^{\i
  th_2}w'\big)\Big)=0.\label{qua9}\eeq
Then $h_1=h_2$.\label{qua8}\eel

\proof
Let $w\in\Dom h_1$ and $w'\in\Dom h_2$. Then 
(\ref{qua9}) implies
\beq(h_1 w|w')=(w|h_2 w').\eeq
This means that $h_1\subset h_2^*=h_2$ and $h_2\subset h_1=h_1^*$. Hence
$h_1=h_2$.
\qed

\end{document}